\documentclass[aps,pre,preprint,showpacs,superscriptaddress,nofootinbib]{revtex4-1}

\usepackage{graphicx}
\usepackage{bm}
\usepackage{eucal}
\usepackage{mathrsfs}
\graphicspath{{./}}
\usepackage{color}

\newcommand {\be} {\begin{equation}}
\newcommand {\ee} {\end{equation}}
\newcommand {\Be}{\begin{eqnarray*}}
\newcommand {\Ee} {\end{eqnarray*}}
\newcommand {\bey} {\begin{eqnarray}}
\newcommand {\eey} {\end{eqnarray}}
\newcommand{\bit}{\begin{itemize}}
\newcommand{\eit}{\end{itemize}}
\newcommand{\bfl}{\begin{flusleft}}
\newcommand{\efl}{\end{flusleft}}
\newcommand{\bfr}{\begin{flushright}}
\newcommand{\bc}{\begin{center}}
\newcommand{\ec}{\end{center}}
\newcommand{\ben}{\begin{enumerate}}
\newcommand{\een}{\end{enumerate}}

\newcommand{\comment}[1]{}

\newcommand{\red}{\textcolor{red}}

\begin{document}


\title{Hysteretic transitions in the Kuramoto model with inertia}



\date{\today}

\author{Simona Olmi}
\affiliation{CNR - Consiglio Nazionale delle Ricerche - Istituto dei Sistemi
Complessi, via Madonna del Piano 10, I-50019 Sesto Fiorentino, Italy}
\affiliation{INFN Sez. Firenze, via Sansone, 1 - I-50019 Sesto Fiorentino, Italy}

\author{Adrian Navas}
\affiliation{Centre for Biomedical Technology (UPM)
28922 Pozuelo de Alarc\'on, Madrid, Spain}

\author{Stefano Boccaletti}
\affiliation{CNR - Consiglio Nazionale delle Ricerche - Istituto dei Sistemi
Complessi, via Madonna del Piano 10, I-50019 Sesto Fiorentino, Italy}
\affiliation{INFN Sez. Firenze, via Sansone, 1 - I-50019 Sesto Fiorentino, Italy}
\affiliation{Centre for Biomedical Technology (UPM)
28922 Pozuelo de Alarc\'on, Madrid, Spain}

\author{Alessandro Torcini}
\affiliation{CNR - Consiglio Nazionale delle Ricerche - Istituto dei Sistemi
Complessi, via Madonna del Piano 10, I-50019 Sesto Fiorentino, Italy}
\affiliation{INFN Sez. Firenze, via Sansone, 1 - I-50019 Sesto Fiorentino, Italy}


\date{\today}

\begin{abstract}
We report finite size numerical investigations and mean field analysis of
a Kuramoto model with inertia for fully coupled and diluted systems.
In particular, we examine for a Gaussian distribution of the frequencies
the transition from incoherence to coherence for increasingly 
large system size and inertia.
For sufficiently large inertia the transition is hysteretic and
within the hysteretic region clusters of locked oscillators of various sizes 
and different levels of synchronization coexist. A modification of 
the mean field theory developed by Tanaka, Lichtenberg, and Oishi [{\it Physica D, 100 (1997) 279}]
allows to derive the synchronization profile associated to each of these clusters.
We have also investigated numerically the \red{limits of existence} of the coherent and of 
the incoherent \red{solutions}. \red{The minimal coupling required to observe the coherent state is
largely independent of the system size and it saturates to a constant value already for moderately
large inertia values. The incoherent state is observable up to a critical coupling
whose value saturates for large inertia and for finite system sizes,  
while in the thermodinamic limit this critical value 
diverges proportionally to the mass.}
By increasing the inertia the transition becomes more complex, and the synchronization
occurs via the emergence of clusters of whirling oscillators. The presence
of these groups of coherently drifting oscillators 
induces oscillations in the order parameter. We have shown that
the transition remains hysteretic even for randomly diluted networks 
up to a level of connectivity corresponding to few links per oscillator.
Finally, an application to the Italian high-voltage power grid is reported,
which reveals the emergence of quasi-periodic oscillations in the order
parameter due to the simultaneous presence of many competing whirling clusters.
\end{abstract}


\pacs{05.45.Xt, 05.45.-a, 64.60.aq, 89.75.-k}


\maketitle

\section{Introduction}
\label{sec1}

 Synchronization phenomena in phase oscillator networks are usally addressed
by considering the paradagmatic Kuramoto model~\cite{kura,strogatz2000,pikovsky2003,acebron2005}.
This model has been applied in many contexts ranging from 
crowd synchrony~\cite{strogatz2005} to synchronization, learning and multistability 
in neuronal systems~\cite{cumin2007,niyogi2009,maistrenko2007}. Furthermore,
the model has been considered with different topologies ranging from homogeneous 
fully coupled networks to scale-free inhomogeneous systems~\cite{arenas2008}. Recently, 
it has been employed as a prototypical example to analyze
low dimensional behaviour in a single large population 
of phase oscillators with a global sinusoidal coupling~\cite{ott2008,marvel2009}, 
as well as in many hierarchically coupled sub-populations~\cite{pikovsky20008}.
The study of the Kuramoto model for non-locally coupled arrays
~\cite{kuramoto2002, abrams2004} and for two populations of symmetrically globally 
coupled oscillators~\cite{abrams2008} lead to the discover of the so-called Chimera states,
whose existence has been revealed also experimentally in the very 
last years~\cite{hagerstrom2012, tinsley2012, martens2013, larger2013}.

In this paper we will examine the dynamics and synchronization properties
of a generalized Kuramoto model for phase oscillators with an inertial term both for fully coupled 
and for diluted systems. The modification of the Kuramoto model with an additional inertial term was 
firstly reported in~\cite{tanaka1997first,tanaka1997self} by Tanaka, Lichtenberg
and Oishi (TLO). These authors have been inspired in their modelization by 
a previous phase model developed by Ermentrout to mimic the synchronization mechanisms
observed among the fireflies {\it Pteroptix Malaccae}~\cite{ermentrout1991}.
These fireflies synchronize their flashing activity by entraining to the forcing
frequency with almost zero phase lag, even for stimulating frequencies different from
their own intrinsic flashing frequency. The main ingredient to allow for 
the adaptation of the flashing frequency to the forcing one is to include
an inertial term in a standard phase model for synchronization.
Furthermore, networks of phase coupled oscillators with inertia have been recently employed 
to investigate the self-synchronization in power grids~\cite{salam1984,filatrella2008,rohden2012},
as well as in disordered arrays of underdamped Josephson junctions~\cite{trees2005}.
Explosive synchronization have been reported for a complex system
made of phase oscillators with inertia, where the natural frequency of each
oscillator is assumed to be proportional to the degree of its node~\cite{Ji2013}.
In particular, the authors have shown that the TLO mean field approach reproduces
very well the numerical results for their system.

Our aim is to describe from a dynamical point of view the hysteretic transition observed in the TLO model
for finite size systems and for various values of the inertia; we will
devote a particular emphasis to the description and characterization of coexisting 
clusters. Furthermore, the analysis is extended to random networks
for different level of dilution and to a realistic case, represented by
the high-voltage power grid in Italy.
In particular, in Sect. \ref{sec2} we will introduce the model and we will describe 
our simulation protocols
as well as the order parameter employed to characterize the level of coherence
in the system. The mean field theory developed by TLO is presented in Sect.~\ref{sec3}
together \red{with a generalization able to capture the emergence of 
clusters of locked oscillators of any size 
induced by the presence of the inertial term}. The theoretical mean field results 
are compared with finite size simulations of fully
coupled systems in Sect.~\ref{sec4}; in the same Section the stability
limits of the coherent and incoherent phase are numerically investigated \red{for various
simulation protocols} as a function of the mass value and of the system size. A last subsection is devoted
to the emergence of clusters of drifting oscillators and to their influence on the
collective level of coherence.  The hysteretic transition for random diluted networks is examined
in the Sect.~\ref{sec5}. \red{As a last point the behaviour of the model is analyzed 
for a network architecture corresponding to the Italian high-voltage power grid in Sect.~\ref{sec6}}. 
Finally, the reported results are briefly summarized and discussed in Sect.~\ref{sec7}.

\section{Simulation Protocols and Coherence Indicators}
\label{sec2}

By following Refs.~\cite{tanaka1997self,tanaka1997first},
we study the following version of the Kuramoto model with inertia:
\red{
\begin{equation}
\label{eqPRL}
 m\ddot{\theta}_i + \dot{\theta}_i=\Omega_i + \frac{K}{N_i}\sum_j
C_{i,j} \sin(\theta_j-\theta_i)
\end{equation}
}
where $\theta_i$ and $\Omega_i$ are, respectively, the instantaneous phase and
the natural frequency of the $i$-th oscillator, $K$ is the coupling,
\red{the matrix element $C_{i,j}$ takes the value one (zero) depending if the
link between oscillator $i$ and $j$ is present (absent) and $N_i$ is the
in-degree of the $i$-th oscillator}. For a fully connected networks $C_{i,j} \equiv 1$
and \red{$N_i = N$}; for the diluted case we have considered \red{undirected
random graphs with a constant in-degree $N_i = N_c$, therefore 
each node has exactly $N_c$ random connections and $C_{i,j}=C_{j,i}$}.
In the following \red{we will mainly consider natural frequencies $\Omega_i$} 
randomly distributed according to
a Gaussian distribution $g(\Omega)=\frac{1}{\sqrt{2\pi}} {\rm e}^{-\Omega^2/2}$ 
with zero average and an unitary standard deviation.

To measure the level of coherence between the oscillators, we
employ the complex order parameter~\cite{winfree}
\begin{equation}
\label{order_parameter}
 r(t)e^{i\phi(t)}=\frac{1}{N}\sum_j e^{i\theta_j} \enskip ;
\end{equation}
where $r(t) \in [0:1]$ is the modulus and $\phi(t)$ the phase of the 
considered indicator. An asynchronous state, in a finite network,
is characterized by $r \simeq \frac{1}{\sqrt{N}}$, while 
for $r \equiv 1$ the oscillators are fully synchronized
and intermediate $r$-values correspond to partial synchronization.
Another relevant indicator for the state of the network is the number of
locked oscillators $N_L$, characterized by the same (vanishingly)
small average phase velocity $<\theta_i>$, and the maximal locking 
frequency $\Omega_M$, which corresponds to the maximal natural frequency
$|\Omega_i|$ of the locked oscillators.
 
In general we will perform sequences of simulations \red{by varying adiabatically
the coupling parameter $K$ with two different protocols}.
Namely, for the first protocol (I) the series of simulations is initialized 
for the decoupled system by considering random initial  conditions for $\{\theta_i\}$ 
and $\{\dot \theta_i\}$. Afterwards the coupling is increased in steps
$\Delta K$ until a maximal coupling $K_M$ is reached. For each value
of $K$, apart the very first one, the simulations is initialized by 
employing the last configuration of the previous simulation in the sequence.
For the second protocol (II), starting from the final coupling $K_M$ achieved
\red{by employing} the protocol (I) simulation, the coupling is reduced in steps $\Delta K$
until $K=0$ is recovered. At each step the system is simulated for a transient time $T_R$ 
followed by a period $T_W$ during which the average value of the order parameter 
${\bar r }$ and  of the velocities $\{ \langle \dot \theta \rangle \}$, as well as $\Omega_M$,
are estimated.

\begin{figure}
\includegraphics[angle=0,width=7.5cm]{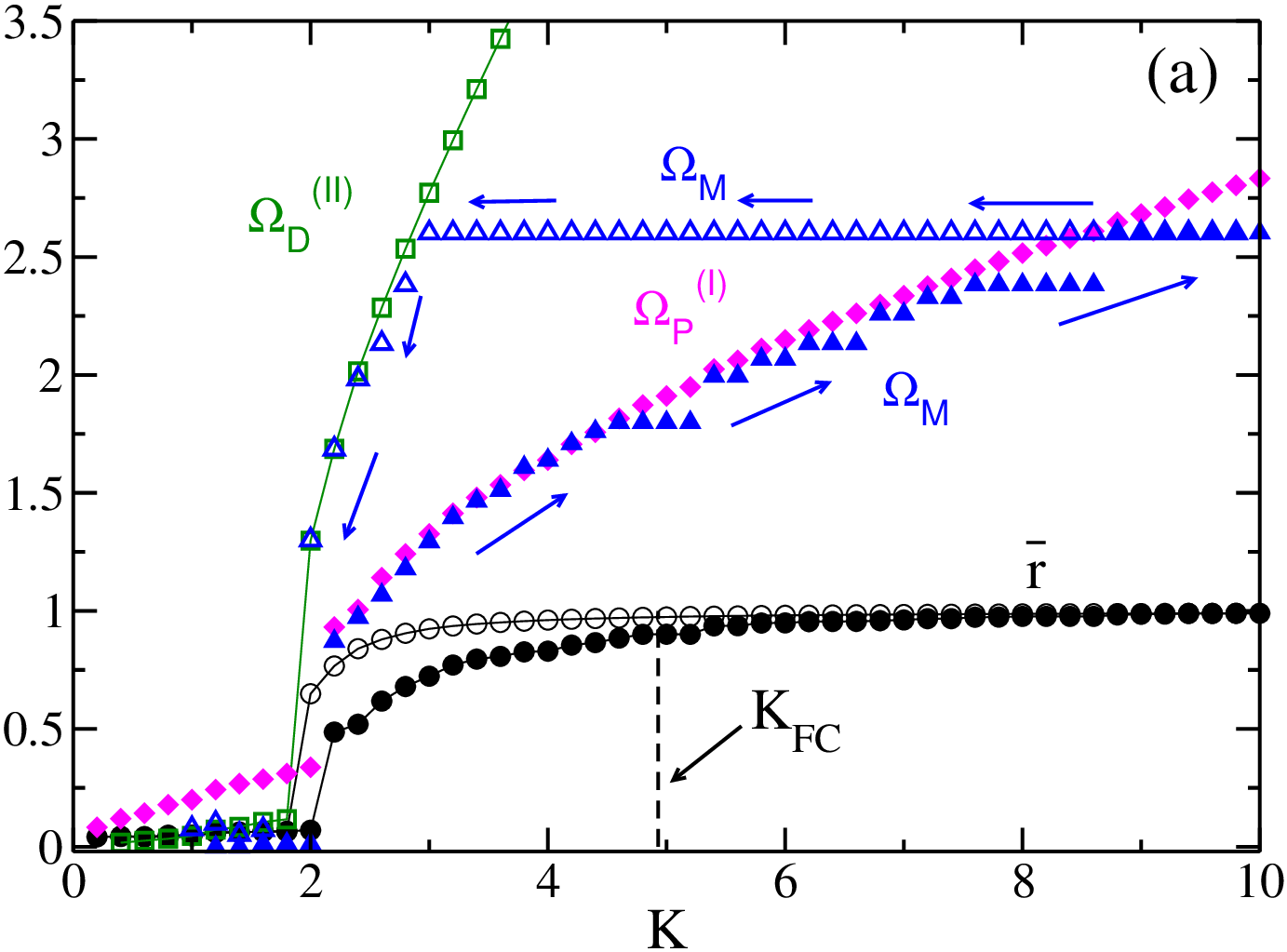}
\includegraphics[angle=0,width=7.5cm]{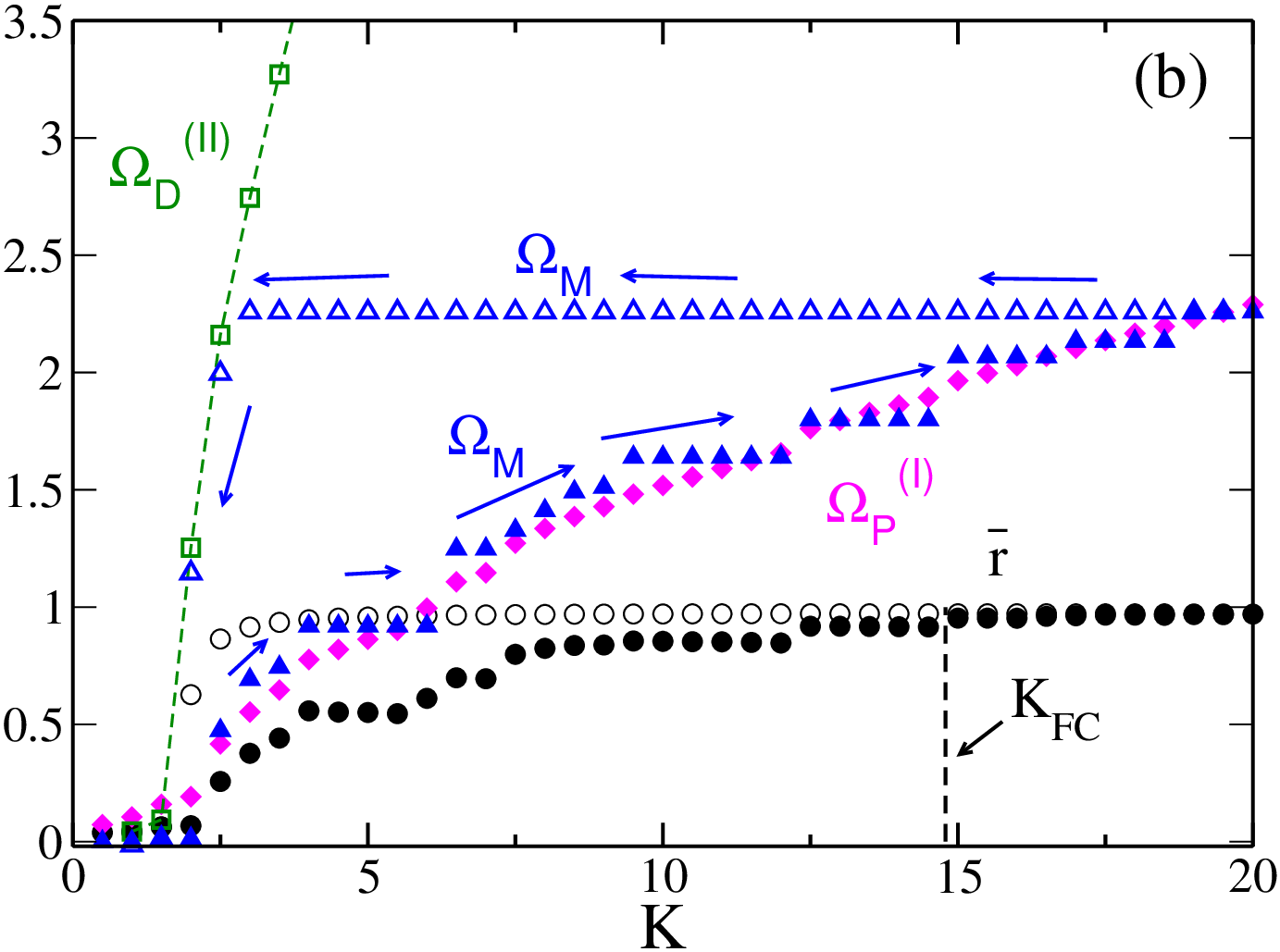}
\caption{(Color Online) Average order parameter ${\bar r }$ (black circles) 
and maximal locking frequency $\Omega_M$ (blue triangles) as a function of the coupling $K$
for two series of simulations performed following the protocol (I)
(filled symbols) and the protocol (II) (empty symbols).
The data refer to mass: $m=2$ (a) and $m=6$ (b).
For $m=2$ ($m=6$) we set $\Delta K=0.2$ ($\Delta K =0.5$)
and $K_M =10$ ($K_M=20$), in both cases $N=500$, $T_R = 5,000$
and $T_W =200$. The (magenta) diamonds indicate $\Omega_P^{(I)} = \frac{4}{\pi} \sqrt{\frac{K \bar r  }{m}}$
for protocol (I), the (green) squares $\Omega_D^{(II)} = K \bar r$ for protocol (II), and the (black) dashed vertical 
line \red{the coupling constant $K^G_{FC}$, whose expression is reported in Eq.~(\ref{Gau_KFC})}.
\label{omegadis}
}
\end{figure}

An example of the outcome obtained by performing the sequence of
simulations of protocol (I) followed by protocol (II) is reported in 
Fig.~\ref{omegadis} for not negligible inertia, namely, $m=2$ and $m=6$.
During the first series of simulations (I) the system remains desynchronized 
up to a threshold $K = K_1^c \simeq 2$, above this value $\bar r $ shows a jump 
to a finite value and then increases with $K$, saturating to $\bar r  \simeq 1$ at sufficiently large 
coupling
\footnote{Please notice that in the data shown in Fig.~\ref{omegadis}
the final state does not correspond to the 100\% of synchronized oscillators, but to
99.6 \% for $m=2$ and 97.8 \% for $m=6$. However, the reported considerations are not
modified by this minor discrepancy.}.
By decresing $K$ one observes that the value of $\bar r $ assumes larger values
than during protocol (I), while the system desynchronizes at
a smaller coupling, namely $K_2^c < K_1^c$.
Therefore, the limit of stability of the asynchronous state is given by $K_1^c$,
while the partially synchronized state can exist down to $K_2^c$, thus asynchronous and
partially synchronous states coexist in the interval $[K_2^c;K_1^c]$.

The maximal locking frequency $\Omega_M$ increases with $K$ during the first phase.
In particular, for sufficiently large coupling, $\Omega_M$ displays plateaus followed 
by jumps for large coupling: this indicates that the oscillators frequencies $\Omega_i$ are
grouped in small clusters. Finally, for $\bar r  \simeq 1$ the frequency
$\Omega_M$ attains a maximal value.  By reducing the coupling, following now
the protocol (II), $\Omega_M$ remains stucked to such a value for a large $K$ interval.
Then $\Omega_M$ reveals a rapid decrease towards zero for small coupling $K \simeq K_2^c$.
In the next Section, we will give an interpretation of this behaviour. 

We will also perform a series of simulations with a different protocol (S),
\red{to test for the independence of the results reported
for $K_1^c$ and $K_2^c$ from the chosen initial conditions.}
In particular, for a certain coupling $K$ we consider an asynchronous initial condition 
and we {\it perturb} such a state by forcing all the neurons with natural frequency 
$|\Omega_i| < \omega_S $ to be locked. Namely, we initially set their velocities and phase
to zero, then we let evolve the system for a transient time $T_R$ followed by a period $T_W$ during which 
${\bar r }$ and the other quantities of interest are measured.  These simulations will
be employed to identify the \red{interval of coupling parameters over which the coherent 
and incoherent solutions can be numerically observed}. 

In more details, to measure with this approach $K_1^c$,
which represents \red{the upper coupling value for which the incoherent state can be observed},
we fix the coupling $K$ and we perform a series of simulations 
for increasing $\omega_S$ values, namely from $\omega_S=0$ to $\omega_S=3$
in steps $\Delta \omega_S =0.05$. \red{For each simulation we measure the order
parameter $\bar r$, whenever it is finite for some $\omega_S > 0$,
the corresponding coupling is associated to a partially synchronized state, the smallest
coupling for which this occurs is identified as $K_c^1$. }

In order to identify $K_2^c$, which is the \red{lower value of the coupling
for which the coherent state is numerically observable}, we 
measure the minimal $K$ for which the unperturbed asynchronous state (corresponding to
$\omega_S=0$) spontaneously evolves towards a partially synchronized solution.
To give a statistically meaningfull estimation of $K_c^1$ and $K_2^c$, we have averaged
the results obtained for various different initial conditions, ranging from 5 to 8,
for all the considered system sizes and masses.

\red{In principle, this approach cannot test rigorously for the stability 
of the coherent and incoherent states, since it deals with a very specific perturbation
of the initial state. However, as we will show the estimations of the critical couplings 
obtained with protocol (S) coincide with those given by protocols (I) and (II),
thus indicating that the reported results are not critically dependent on the chosen
initial conditions.
}

\begin{figure}
\includegraphics[angle=0,width=7.5cm]{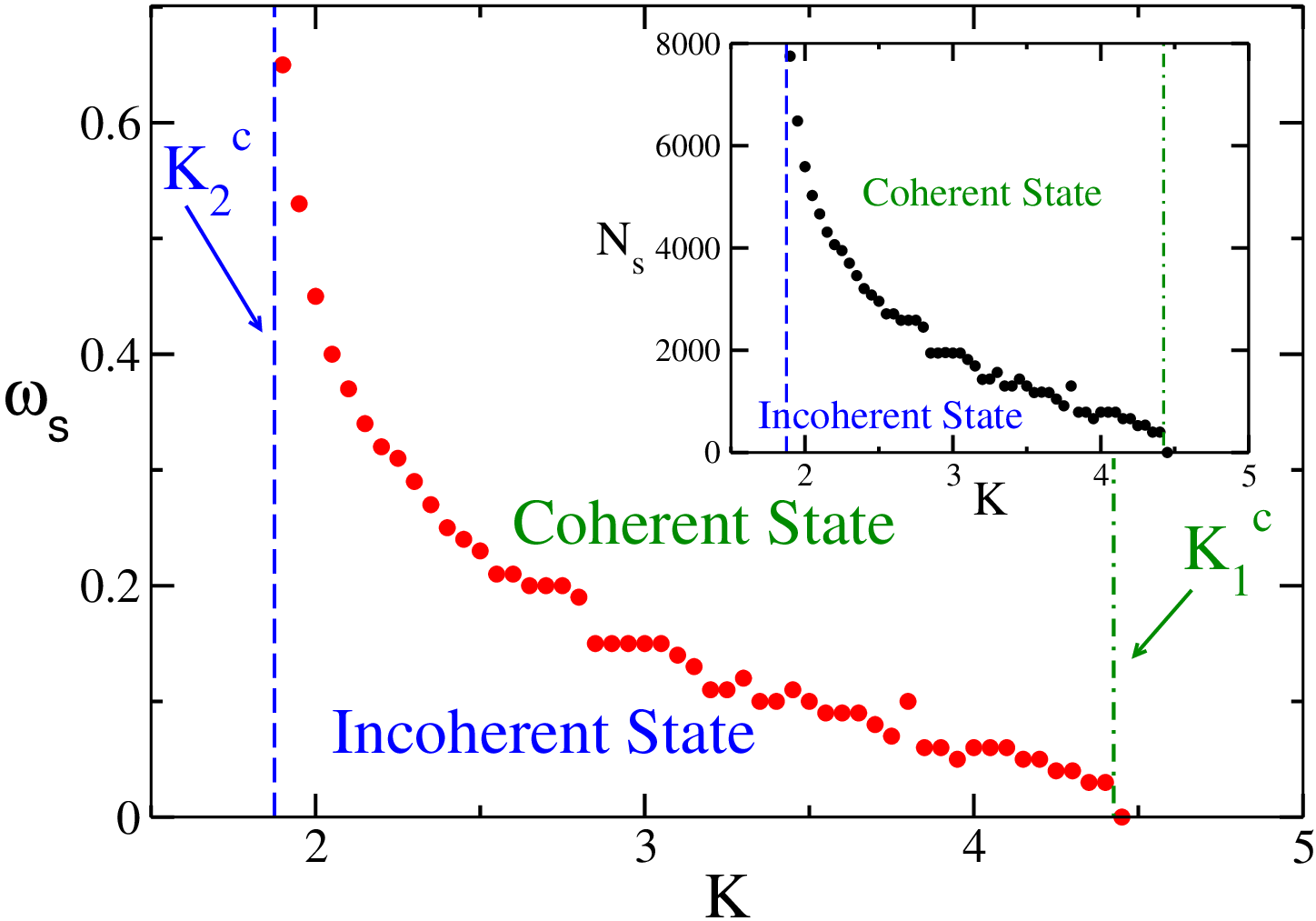}
\caption{(Color Online) Minimal $\omega_S$ giving rise to a state characterized by a finite
level of synchronization (i.e. $\bar r > 0$) as a function of the coupling constant $K$.
The inset reports \red{the minimal number $N_S$ of oscillators which should be initially locked
in order to lead to the emergence of a coherent state,} as a function of $K$.
The vertical (green) dot-dashed line refers to the estimated $K_1^c$ and the 
(blue) dashed line indicates the estimated $K_2^c$.
The data refer to simulations performed with protocol (S) for $N =16,000$,
$m=6$, with $T_W=2,000$ and $T_R=20,000$. 
\label{prot3}
}
\end{figure}

\section{Mean Field Theory}
\label{sec3}

In the fully coupled case Eq. (\ref{eqPRL}) can be rewritten, by employing the
order parameter definition (\ref{order_parameter}) as follows
\begin{equation}
\label{eqRED}
 m\ddot{\theta}_i + \dot{\theta}_i=\Omega_i -  K r \sin(\theta_i -\phi)
\qquad ;
\end{equation}
which corresponds to a damped driven pendulum equation. This equation admits
for sufficiently small forcing frequency $\Omega_i$ two fixed points: 
a stable node and a saddle. At larger frequencies $\Omega_i > \Omega_P 
\simeq \frac{4}{\pi}\sqrt{\frac{Kr}{m}}$ a homoclinic bifurcation
leads to the emergence of a limit cycle from the saddle. The stable
limit cycle and the stable fixed point coexist until a saddle node 
bifurcation, taking place at $\Omega_i = \Omega_D = K r$, leads to the disappearence
of the fixed points and for  $\Omega_i  > \Omega_D$ only the oscillating solution
is presents. This scenario is correct for sufficiently large masses,
at small $m$ one have a direct transition from a stable node to
a periodic oscillating orbit at $\Omega_i = \Omega_D = K r$~\cite{strogatz}.

 Therefore for sufficiently large $m$ there is a coexistence regime
where, depending on the initial conditions, the single oscillator 
can rotate or stay quiet. How this single unit property will reflect in the
self-consistent collective dynamics of the coupled systems is the topic
of this paper.

\red{
\subsection{The Theory of Tanaka, Lichtenberg, and Oishi}
}

Tanaka, Lichtenberg, and Oishi in their seminal papers \cite{tanaka1997self,tanaka1997first}
have examined the origin of the first order hysteretic transition observed
for Lorentzian and flat (bounded) frequency distributions $g(\Omega)$
by considering two different initial states for the network :
(I) the completely desynchronized state ($r=0$) and (II) the fully
synchronized one ($r \equiv 1$). Furthermore, in case I (II) they studied
how the level of synchronization, measured by $r$, varies due to the 
increase (decrease) of the coupling $K$. In the first case the oscillators
are all initially drifting with finite velocities $\langle \dot \theta_i \rangle$; 
by increasing $K$ the oscillators with smaller 
natural frequencies $|\Omega_i| < \Omega_P$ begin to lock
($\langle \dot \theta_i \rangle=0$), while the other continue to drift.
This picture is confirmed by the data reported in Fig.~\ref{omegadis}, where
the maximal value $\Omega_M$ of the frequencies of the locked oscillators is
well approximated by $\Omega_P$. \red{The process continues until all the oscillators
are finally locked leading to $r =1$.}

In the case (II), TLO assumed that initially all the oscillators were already locked,
with an associated order parameter $r \equiv 1$. Therefore, the oscillators can start
to drift only when the stable fixed point solution will disappear, leaving the system 
only with the limit cycle solution. This happens, by decreasing $K$,
whenever $|\Omega_i| \ge \Omega_D = K r$. This is numerically verified,
indeed, as shown in Fig.~\ref{omegadis},
the maximal locked frequency $\Omega_M$ remains constant
until, by decreasing $K$, it encounters the curve $\Omega_D$ and then
$\Omega_M$ follows this latter curve down to the desynchronized state.
The case (II) corresponds to the situation observable for the usual Kuramoto model, 
where there is no bistability~\cite{kura}.

In both the examined cases there is a group of desynchronized oscillators
and one of locked oscillators separated by a frequency, $\Omega_P$ in the first
case and $\Omega_D$ in the second one. These groups contribute differently to the total level of 
synchronization of the system, namely
 \begin{equation}
\label{rrr}
r = r_L + r_D
\end{equation}
where $r_L$ ($r_D$) is the contribution of the locked (drifting) population.

For the locked population, one gets
\begin{equation}
\label{rL}
r_L^{I,II} = K r \int_{-\theta_{P,D}}^{\theta_{P,D}} \cos^2 \theta g(K r \sin \theta) d \theta
\qquad ;
\end{equation}
where $\theta_P = \sin^{-1} (\Omega_P/Kr)$ and $\theta_D =\sin^{-1} (\Omega_D/Kr) \equiv \pi/2$.

The drifting oscillators contribute to the total order parameter with
a negative contribution; the self-consistent integral defining $r_D$
\red{has been estimated by TLO} in a perturbative manner by performing an expansion up
to the fourth order in $1/(m K)$ and $1/(m \Omega)$. 
\red{
Therefore the obtained expression is correct in the limit of sufficiently
large masses and it reads as
\begin{equation}
\label{rD}
r_D^{I,II} \simeq - m K r \int_{\Omega_{P,D}}^{\infty}  \frac{1}{(m \Omega)^3} g(\Omega) d \Omega
\qquad ;
\end{equation}
where $g(\Omega) = g(-\Omega)$.}

By considering an initially desynchronized (fully synchronized) system 
and by increasing (decreasing) $K$ one can get a theoretical approximation 
for the level of synchronization in the system by employing the 
mean-field expression (\ref{rL}), (\ref{rD}) and (\ref{rrr}) for case I (II). 
In this way, two curves are obtained in the phase plane $(K,r)$, namely
$r^I (K)$ and $r^{II} (K)$. In the following, we will show that these are not 
the unique admissible solutions in the 
mentioned plane, and these curves represent the lower and upper bound
for the possible states characterized by a partial level of synchronization.

Let us notice that the expression for $r_L$ and $r_D$
reported in Eqs. (\ref{rL}) and (\ref{rD}) are the same 
for case (I) and (II), only the integration
extrema have been changed. These are defined by the frequency which
discriminates locked from drifting neuron, that
in case (I) is $\Omega_P$ and in case (II) $\Omega_D$.
The value of these frequencies is a function
of the order parameter $r$ and of the coupling constant $K$,
therefore by increasing (decreasing) $K$ they change accordingly.

However, in principle one could fix the discriminating frequency to some
arbitrary value $\Omega_0$ and solve self-consistently the equations
Eqs. (\ref{rrr}), (\ref{rL}), and (\ref{rD}) for 
different values of the coupling $K$. This amounts to solve the following equation
\begin{equation}
\label{r0}
 \int_{-\theta_{0}}^{\theta_{0}} \cos^2 \theta g(K r^0 \sin \theta) d \theta  
- m \int_{\Omega_{0}}^{\infty}  \frac{1}{(m \Omega)^3} g(\Omega) d \Omega = \frac{1}{K}
\qquad ;
\end{equation}
with $\theta_0 = \sin^{-1} (\Omega_0/Kr^0)$. Thus obtaining
a solution $r^0 = r^0(K,\Omega_0)$, which exists provided
that $\Omega_0 \le \Omega_D(K) = r^0 K$. \red{Therefore a portion of the
$(K,r)$ plane, delimited by the curve $r^{II}(K)$,
will be filled with the curves $r_0(K)$ obtained for different
$\Omega_0$ values (as shown in Fig.~\ref{hysteresis} for fully coupled systems and 
in Fig.~\ref {hysteresis_diluita} for diluted ones.).}
\red{These solutions represent clusters of $N_L$ oscillators for which the maximal 
locking frequency and $N_L$ do not vary upon changing the coupling strength. These states 
will be the subject of numerical investigation of the next Sections.
In particular, we will show via numerical simulations that for $K > K_2^c$ these states are 
numerically observables within the portion of the phase space delimited by the two curves 
$r^{I}(K)$ and $r^{II}(K)$ (see Fig.~\ref{hysteresis} and  Fig.~\ref {hysteresis_diluita}).
}

\red{
\subsection{Linear Stability Limit for the Incoherent Solution}
}

As a final aspect, we will report the results of a recent theoretical mean field
approach based on the Kramers description of the evolution of the 
single oscillator distributions for coupled oscillators with inertia and noise~\cite{acebron2000,gupta2014}. 
\red{In particular, the authors in~\cite{gupta2014} have derived  an analytic expression 
for the coupling $K_1^{MF}$, which delimits the range of linear stability for the
asynchronous state. In the limit of zero noise, $K_1^{MF}$
can be obtained by solving the following equation
\begin{equation}
\label{K1c}
\frac{1}{K_1^{MF}} = \frac{\pi g(0)}{2}  - \frac{m}{2} 
 \int_{-\infty}^{\infty} \frac{g(\Omega) d\Omega}{1 + m^2 \Omega^2}
\qquad ;
\end{equation}
where $g(\Omega)$ is an unimodal distribution of width $\sigma$. 
In the limit $m \to 0$ one recovers the value of the critical
coupling for the usual Kuramoto model~\cite{kura}, namely $K_1^{MF} (m\equiv0) = 2/(\pi g(0))$. 
For a Lorentzian distribution an explicit espression for any value of the mass 
can be obtained
\begin{equation}
\label{K1c_lor}
K_1^{MF} = 2 \sigma (1 + m \sigma)
\quad ;
\end{equation}
which coincides with the one reported by Acebr\'on et al~\cite{acebron2000}. 
For a Gaussian distribution it is not possible to find an explicit expression for any $m$, 
however one can derive the first corrective terms to the zero mass limit, namely  
\begin{equation}
\label{K1c_gaumpic}
K_1^{MF} = 2 \sigma \sqrt{\frac{2}{\pi}}\left\{ 1+\sqrt{\frac{2}{\pi}} m \sigma 
+  \frac{2}{\pi} m^2 \sigma^2 +\sqrt{\left(\frac{2}{\pi}\right)^3 -\frac{2}{\pi} } m^3 \sigma^3 \right\}
+ {\cal O}(m^4\sigma^4) \quad .
\end{equation}
On the opposite limit one can analytically show that 
the critical coupling diverges as 
\begin{equation}
\label{K1c_gauminf}
K_1^{MF} \propto 2 m \sigma^2 \qquad  {\rm for} \qquad m \sigma \to \infty \quad.
\end{equation}
It can be seen that this scaling is already valid for not too large masses, 
indeed the analytical results obtained via Eq. (\ref{K1c}) are very well 
approximated, in the range $m \in [1:30]$, by the following expression
\begin{equation}
\label{K1c_gau}
K_1^{MF} \simeq 2 \sigma (0.64 + m \sigma)
\quad .
\end{equation}
This result, together with Eq. (\ref{K1c_lor}), indicates that for both the 
Lorentzian and the Gaussian distribution the critical coupling diverges linearly 
with the mass and quadratically with the width of the frequency distribution.
} 

In the next Section we will compare our numerical results for various system sizes 
with the mean-field result (\ref{K1c}).

\red{
\subsection{Limit of Complete Synchronization}
}

\red{
Complete synchronization can be achieved,  
in the ideal case of infinite oscillators with a distribution $g(\Omega)$ with
infinite support, only in the limit of infinite coupling.}
\red{
However, in finite systems an (almost) complete synchronization is attainable
already at finite coupling, to give an estimation of this effective
coupling $K_{FC}$ one can proceed as follows.
Let us estimate the pinning frequency ${\bar \Omega}_P$ required to have a 
large percentage of oscillators locked, this can be implicitely defined as,
e.g.
\begin{equation}
\label{KFC}
 \int_{-{\bar \Omega}_P}^{{\bar \Omega}_P} g(\Omega) d \Omega = 0.954
\qquad ;
\end{equation}
where by assuming $r \simeq 1$ one sets
${\bar \Omega}_P \simeq \frac{4}{\pi}\sqrt{\frac{K_{FC}}{m}}$ 
and from Eq. (\ref{KFC}) one can derive the coupling $K_{FC}$.
For a Gaussian distribution the integral reported in Eq. (\ref{KFC}) amounts to consider
two standard deviations, and therefore one gets
\begin{equation}
\label{Gau_KFC}
 K_{FC}^G \simeq \left(\frac{\pi}{2}\right)^2 m \sigma^2
\qquad ;
\end{equation}
while for a Lorentzian distribution $g(\Omega) = \frac{\sigma}{\pi} \frac{1}{\sigma^2 +\Omega^2}$
this corresponds to
\begin{equation}
\label{Lor_KFC}
 K_{FC}^L \simeq \left(\frac{13.815 \pi}{4}\right)^2 m \sigma^2
\qquad .
\end{equation}
These results reveal that for increasing mass and width of the frequency 
distribution the system becomes harder and harder to fully synchronize
and that to achieve the same level of synchronization 
a much larger coupling is required for the Lorentzian distribution
(for the same $m$ and $\sigma$).
}

\section{Fully Coupled System}
\label{sec4}

\begin{figure}
\includegraphics[angle=0,width=8.cm]{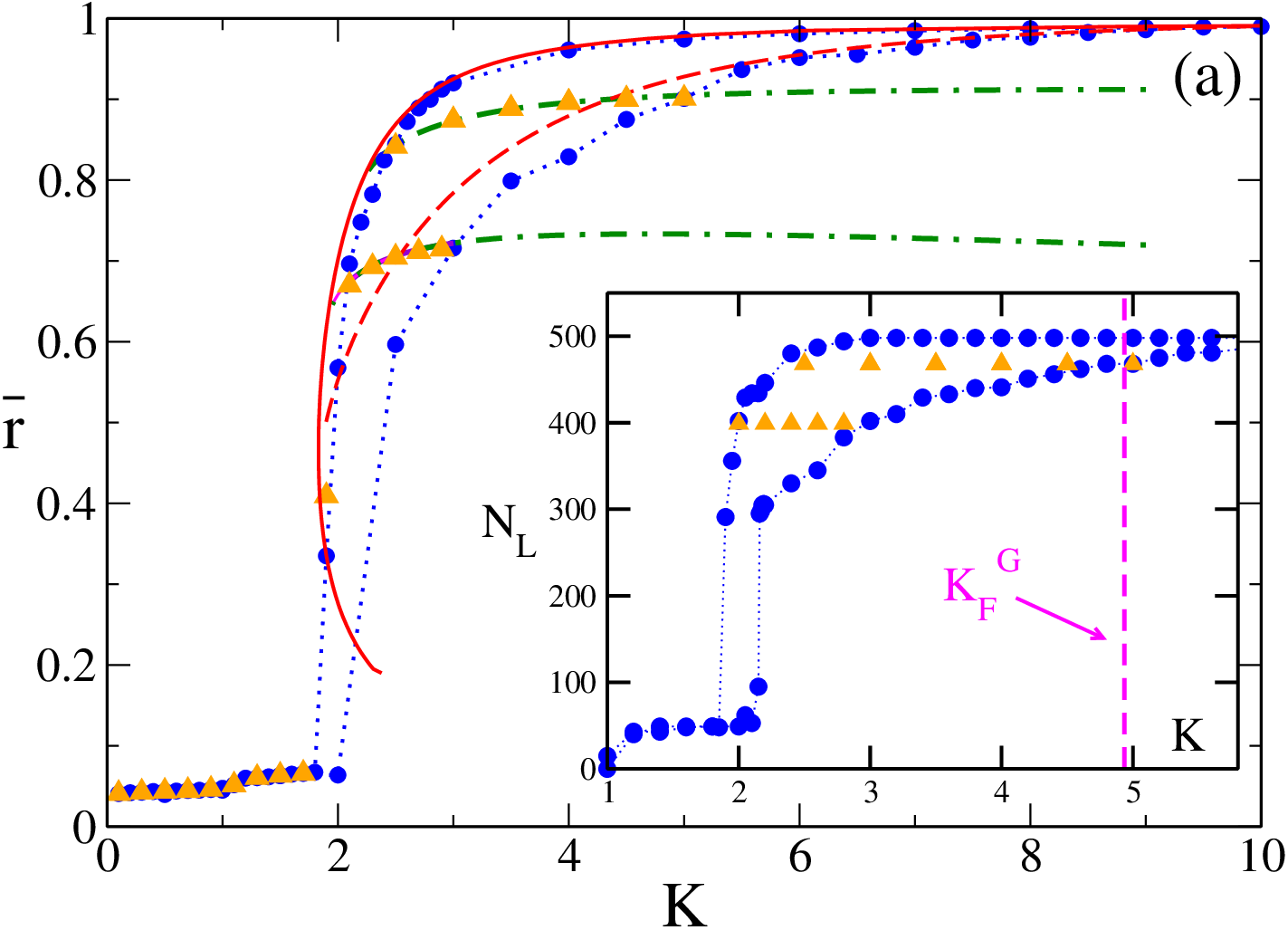}
\includegraphics[angle=0,width=8.cm]{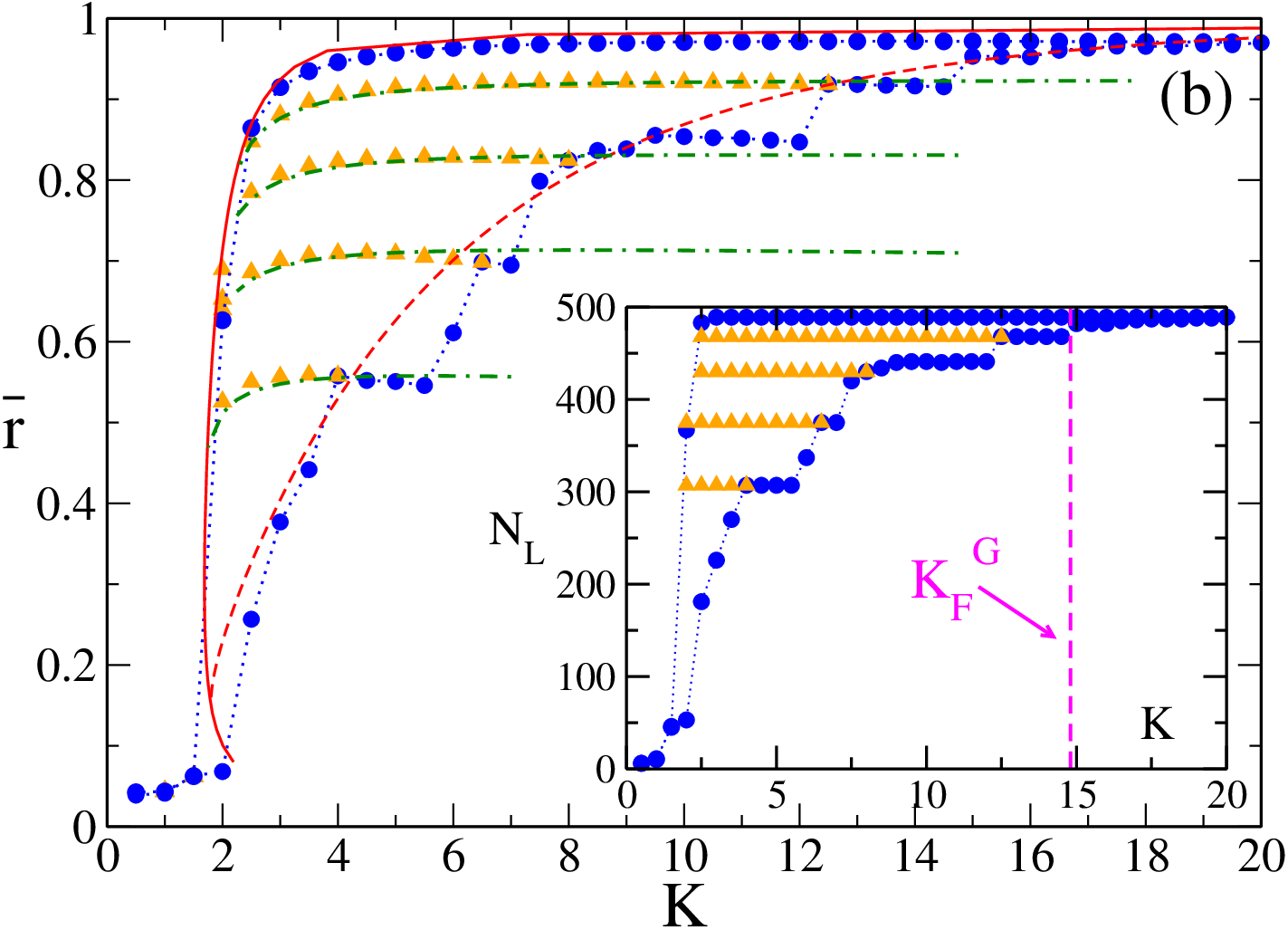}
\caption{(Color Online) Average order parameter ${\bar r}$ versus the coupling constant $K$
for $m=2$ (a) and $m=6$ (b). Mean field estimates: the dashed (solid) red curves refer 
to $r^I  = r^I_L + r^I_D$ ($r^{II}  = r^{II}_L + r^{II}_D$) as obtained by employing
Eqs. (\ref{rL}) and (\ref{rD}) following protocol (I) (protocol (II)); the (green) dot-dashed curves 
are the solutions $r^0(K,\Omega_0)$ of Eq. (\ref{r0}) for different $\Omega_0$ values.
The employed values from bottom to top are:  $\Omega_0=1.21$ and 1.71 in (a) and $\Omega_0=0.79$, 
1.09, 1.31, and 1.79 in (b).Numerical simulations: (blue) filled circles have
been obtained by following protocol (I) and then (II) starting from $K=0$ until
$K_M =10$ ($K_M =20$) for mass $m=2$ ($m=6$) with steps $\Delta K=0.2$ ($\Delta
K = 0.5$); (orange) filled triangles refer to simulations performed by starting
from a final configuration obtained during protocol (I) and by decreasing the coupling
from such initial configurations. The insets display $N_L$ vs $K$ for the numerical simulations
reported in the main figures, the value of $K_F^G$ (eq.(\ref{Gau_KFC})) is also reported 
in the two cases. The numerical data refer to $N=500$, $T_R=5000$, and $T_W=200$.
\label{hysteresis}}
\end{figure}

 In this Section we will compare the analytical results with finite $N$ simulations
for the fully coupled system: a first comparison is reported in Fig.~\ref{hysteresis}
for two different masses, namely $m=2$ and $m=6$. We observe that the 
data obtained by employing the procedure (II) are quite well reproduced from
the mean field approximation $r^{II}$ for both masses (solid red curve in Fig.~\ref{hysteresis}).
This is not the case for the theoretical estimation $r^I$ (dashed red curve), which
for $m=2$ is larger than the numerical data up to quite large coupling, namely $K \simeq 5$;
while for $m=6$,  a better agreement is observable at smaller $K$, however
now $\bar r$ reveals a step-wise structure for the data corresponding to procol (I).
This step-wise structure at large masses is due to the break down of the independence of the 
whirling oscillators: namely, to the formation of locked clusters at non zero velocities~\cite{tanaka1997first}.
Therefore, oscillators join in small groups to the locked solution and not individually as it happens for 
smaller masses; this is clearly revealed by the behaviour of $N_L$ versus the coupling $K$ 
as reported in the insets of Fig.~\ref{hysteresis}(b).

\subsection{Hysteretic Behaviour}
\label{sec4.1}

 As already mentioned, we would like to better investigate the nature of the hysteresis observed
by performing simulations accordingly to protocol (I) or protocol (II).  In particular, we consider
as initial condition a partially synchronized state obtained during protocol (I) for a certain coupling $K_S > K_1$,
then we perform a sequence of consecutive simulations by reducing the coupling at regular steps $\Delta K$.
Some example of the obtained results are shown in Fig.~\ref{hysteresis}, where we report
$\bar r$ and $N_L$ measured during such simulations as a function of the coupling (orange filled triangles).
From the simulations it is evident that the number of locked oscillators $N_L$ remains constant
until we do not reach the descending curve obtained with protocol (II). On the other hand $\bar r$ 
decreases slightly with $K$, this decrease can be well approximated by the mean field solutions 
of Eq. (\ref{r0}), namely $r^0(K,\Omega_0)$ 
with $\Omega _0 = \Omega_P(K_s,r^I(K_S)) = \frac{4}{\pi} \sqrt{\frac{K_s r^I}{m}}$, see the green dot-dashed 
lines in Fig.~\ref{hysteresis} for $m=2$ and $m=6$.
However, as soon as, by decreasing $K$, the frequency $\Omega_0$ becomes equal or smaller than $\Omega_D$,
the order parameter has a rapid drop towards zero following the upper limit curve $r^{II}$.

\begin{figure}
\includegraphics[angle=0,width=8.cm]{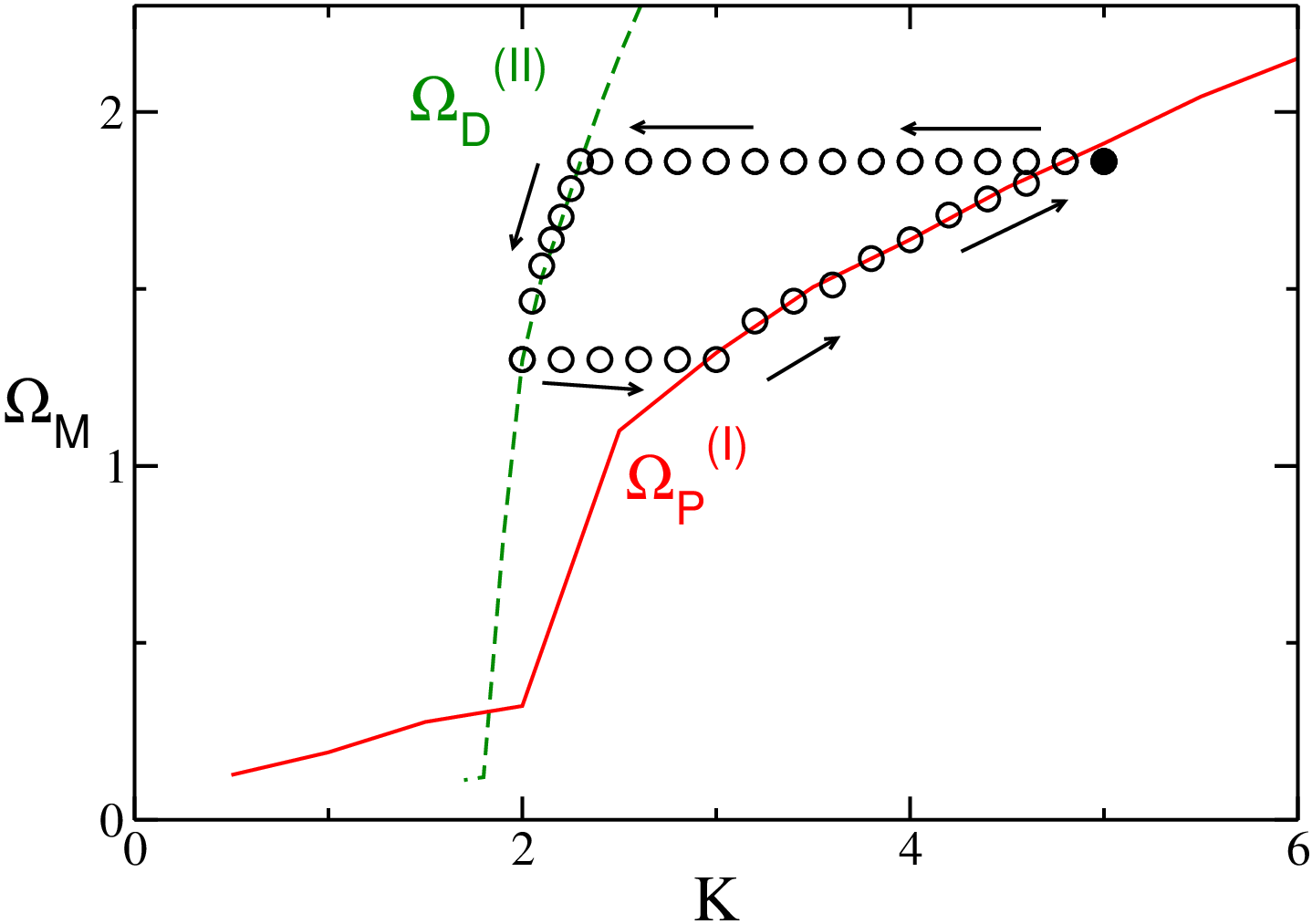}
\caption{(Color Online) Maximal locking frequency $\Omega_M$ versus the coupling
constant $K$. The initial state is denoted by the filled circle at $K_I=5$.
The solid (red) curve indicates the frequency $\Omega_P^{(I)}$ and
the dashed (green) curve the frequency $\Omega_D^{(II)}$.
The numerical data refer to $N=500$, $T_R=5000$,  $T_W=200$, $m=2$, and $\Delta K = 0.1 - 0.05$.
\label{loop}}
\end{figure}

To better interpret these results, let us focus on a simple numerical experiment. We consider
a partially synchronized state obtained for $K_I=5$ with $N=500$ oscillators, then we first decrease
the coupling in steps $\Delta K$ up to a coupling $K_F=2$ and then we increase again $K$ to return
to the initial value $K_I$. During such cyclic simulation we measure $\Omega_M$ for each examined
states, the results are reported in Fig.~\ref{loop}. It is clear that initially
$\Omega_M$ does not vary and it remains identical to its initial value at $K_I=5$.
Furthermore, also the number of locked oscillators $N_L$ remains constant.
The maximal locking frequency (as well as $N_L$) starts to decrease with $K$ only 
after $\Omega_M$ has reached the curve $\Omega_D^{(II)}$, then it follows exactly this curve,  
corresponding to protocol (II), until $K=K_F$.
At this point  we increase again the coupling: the measured $\Omega_M$ 
stays constant at the value $\Omega_D^{(II)} = 2*r^{II}(K_F)$. The frequency
$\Omega_M$ starts to increase only after its encounter with the curve $\Omega_P^{(I)}(K)$.
In the final part of the simulation $\Omega_M$ recovers its initial value by following 
this latter curve. From these simulations it is clear that a synchronized cluster can
be modified by varying the coupling, only by following protocol (I) or protocol (II),
otherwise the coupling seems not to have any relevant effect on the cluster itself.
In other words, all the states $(K,\Omega_M)$ contained between the 
curves $\Omega_D^{(II)}$ and $\Omega_P^{(I)}$ \red{are reachable} 
for the system dynamics, however they are quite peculiar. 

We have verified that the path connecting the initial state at $K_I$ 
to the curve $\Omega_D^{(II)}(K)$, as well as the one connecting 
$K_F$ to the curve $\Omega_P^{(I)}(K)$, are completely reversible.
We can increase (decrease) the coupling from $K_I$ ($K_F$) up to any
intermediate coupling value in steps of any size $\Delta K$ and then decrease (increase) the coupling
to return to $K_I$ ($K_F$) by performing the same steps and the system will pass exactly from the same
states, characterized for each examined $K$ by the values of $\bar r$ and $\Omega_M$.  
Furthermore, as mentioned, there is no dependence on the employed step $\Delta K$, apart the 
restriction that the reached states should be contained within the phase space portion delimited 
by the two curves $\Omega_D^{(II)}$ and $\Omega_P^{(I)}$. 
As soon as the coupling variations would eventually lead the system outside this
portion of the phase space, one should follow a hysteretic loop to return to 
the initial state, similar to the one reported in Fig.~\ref{loop}. 
Therefore, we can affirm that hysteretic loop of any size are possible
within this region of the phase space. \red{For what concerns the stability
of these states, we  can only affirm that from a numerical point of view they appear to
be stable within the considered integration times. However, a (linear) stability
analysis of these solutions is required to confirm our numerical observations.
}

\subsection{Finite Size Effects}
\label{sec4.2}

Let us now examine the influence of the system size on the studied transitions,
in particular we will estimate the transition points $K_1^c$ ($K_2^c$) by considering 
either a sequence of simulations obtained accordingly to protocol (I) (protocol (II))
or asynchronous (synchronous) initial conditions
and by averaging over different realizations of the distributions of the forcing
frequencies $\{\Omega_i\}$.

\begin{figure}
\includegraphics[angle=0,width=8.cm]{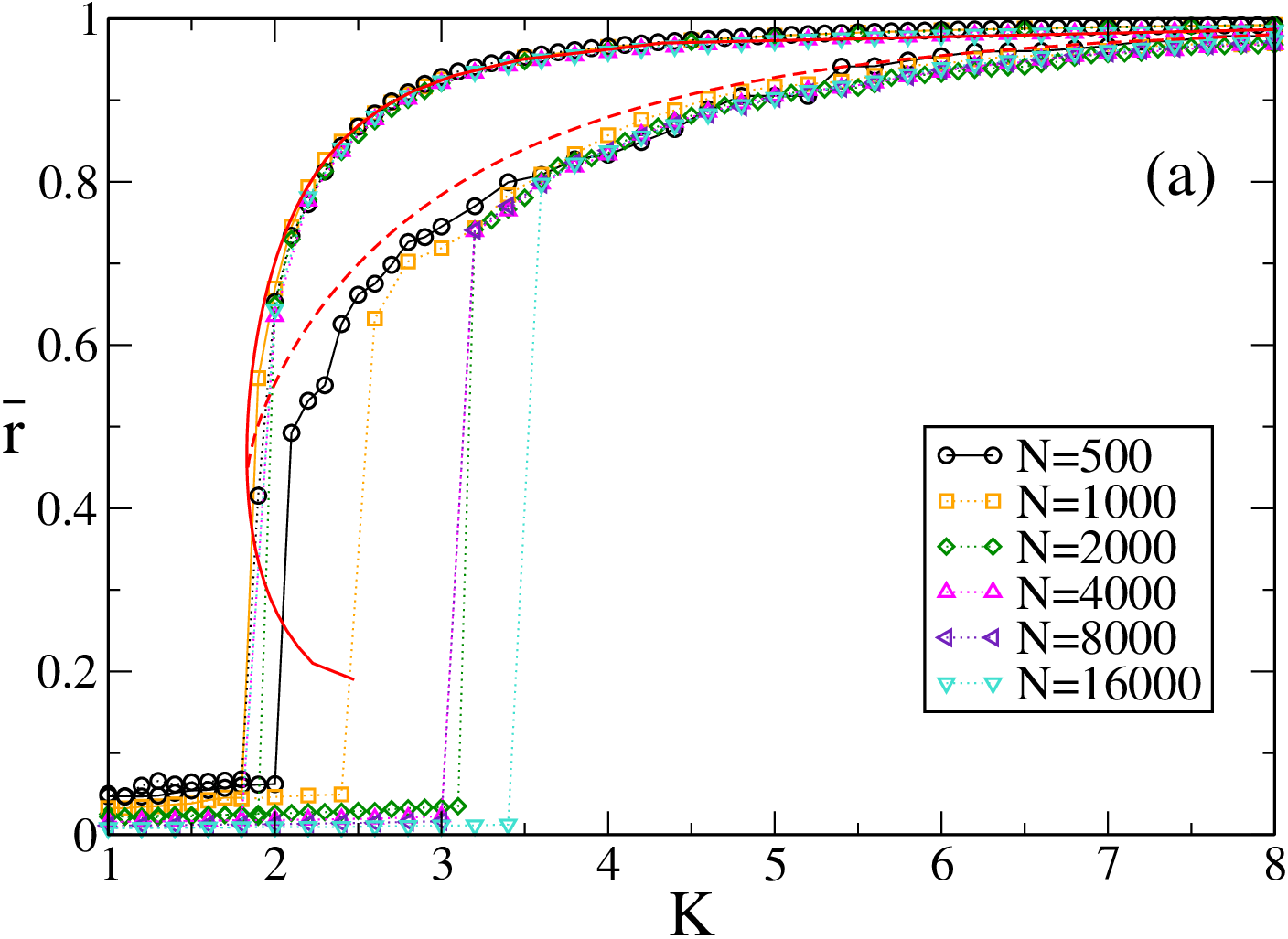}
\includegraphics[angle=0,width=8.cm]{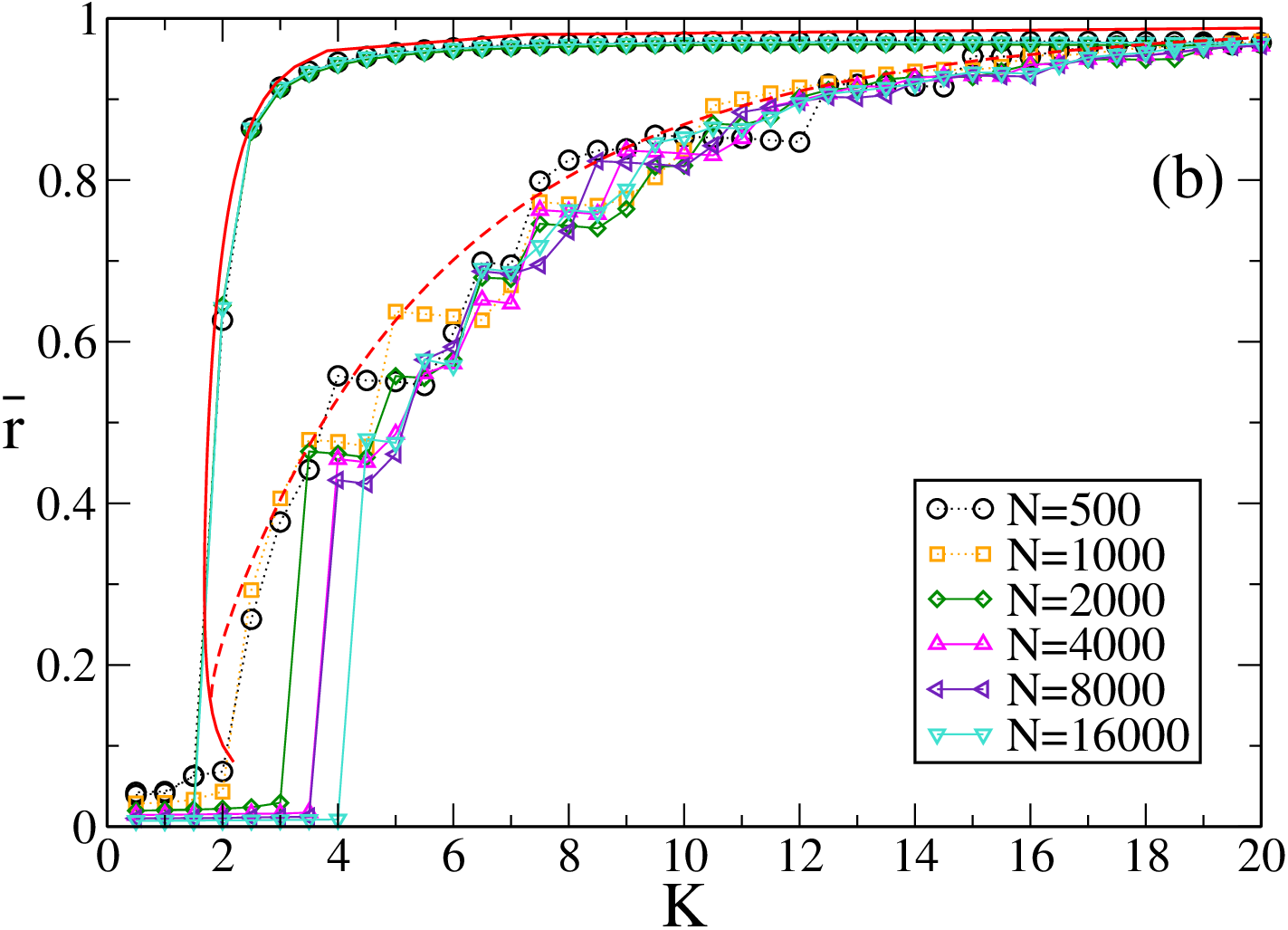}
\caption{(Color Online) Average order parameter $\bar r$  versus the coupling
constant $K$ for various system sizes $N$: (a) $m=2$ and (b) $m=6$.
The (red) solid and dashed curves are the theoretical estimates already reported
in Fig.~\ref{hysteresis}.
The numerical data have been obtained by following
protocol (I) and then protocol (II) from $K=0$ up to $K_M=10$ ($K_M=20$) for
mass $m=2$ ($m=6$) with $\Delta K = 0.2$ ($\Delta K = 0.5$). Data 
have been obtained by averaging the order parameter over a time window $T_W=200$,
after discarding a transient time $T_R \simeq 1,000 - 80,000$ depending on the
system size, the larger $T_R$ have been employed for the larger $N$.
\label{finiteN}}
\end{figure}

The results for the protocol (I) , protocol (II) simulations are
reported in Fig.~\ref{finiteN} for sizes ranging from $N=500$ up to $N=16,000$.
It is immediately evident that $K_2^c$ does not depend heavily on $N$,
while the value of $K_1^c$ is strongly influenced by the size of the
system. Starting from the asynchronous state the system synchronizes
at larger and larger coupling $K_1^c$ with an associated jump in the
order parameter which increases with $N$. 
Whenever the system starts to synchronize, then it follows reasonably well
the mean field TLO prediction and this is particularly true on the way back
towards the asynchronous state along the path associated to protocol (II) 
procedure. \red{However, TLO theory largely fails in giving an estimation
of $K_1^c$ for large system sizes, as shown in Fig.~\ref{finiteN}.}

In the following, we will analyze if the reported finite size results, and in particular the 
values of the critical couplings $K_1^c$ and $K_2^c$, depend on the initial conditions
and on the simulation protocols. For this analysis we focus on two masses, namely $m=2$ and $m=6$,
and we consider system sizes ranging from $N=500$ to $N=16,000$. For each size
and mass we evaluate $K_1^c$ ($K_2^c$) by following protocol (I) (protocol (II)),
as already shown in Fig.~\ref{finiteN}; furthermore now the critical coupling are
also estimated by considering random initial conditions and by applying the protocol (S).

\begin{figure}
\includegraphics[angle=0,width=8.cm]{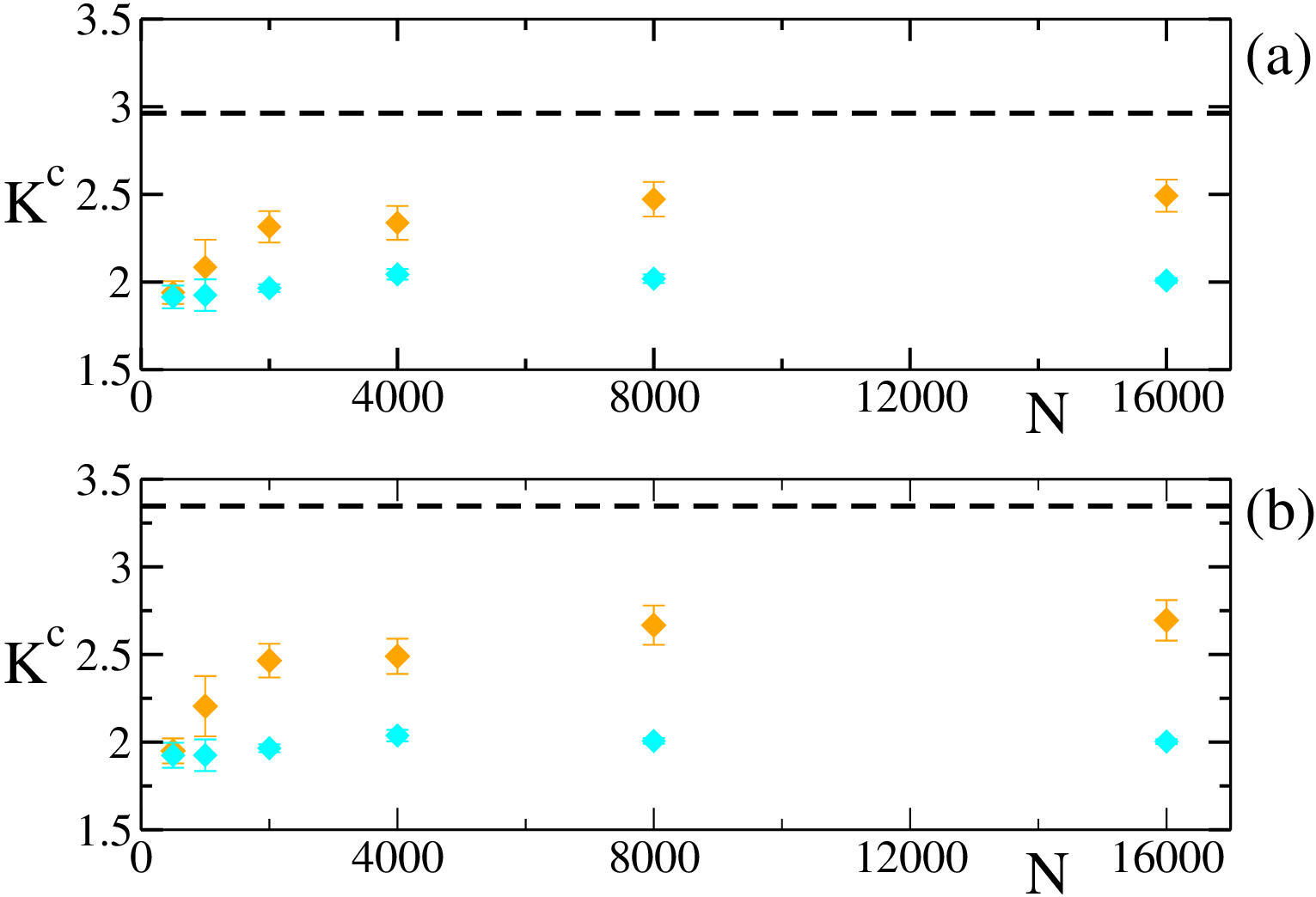}
\includegraphics[angle=0,width=8.cm]{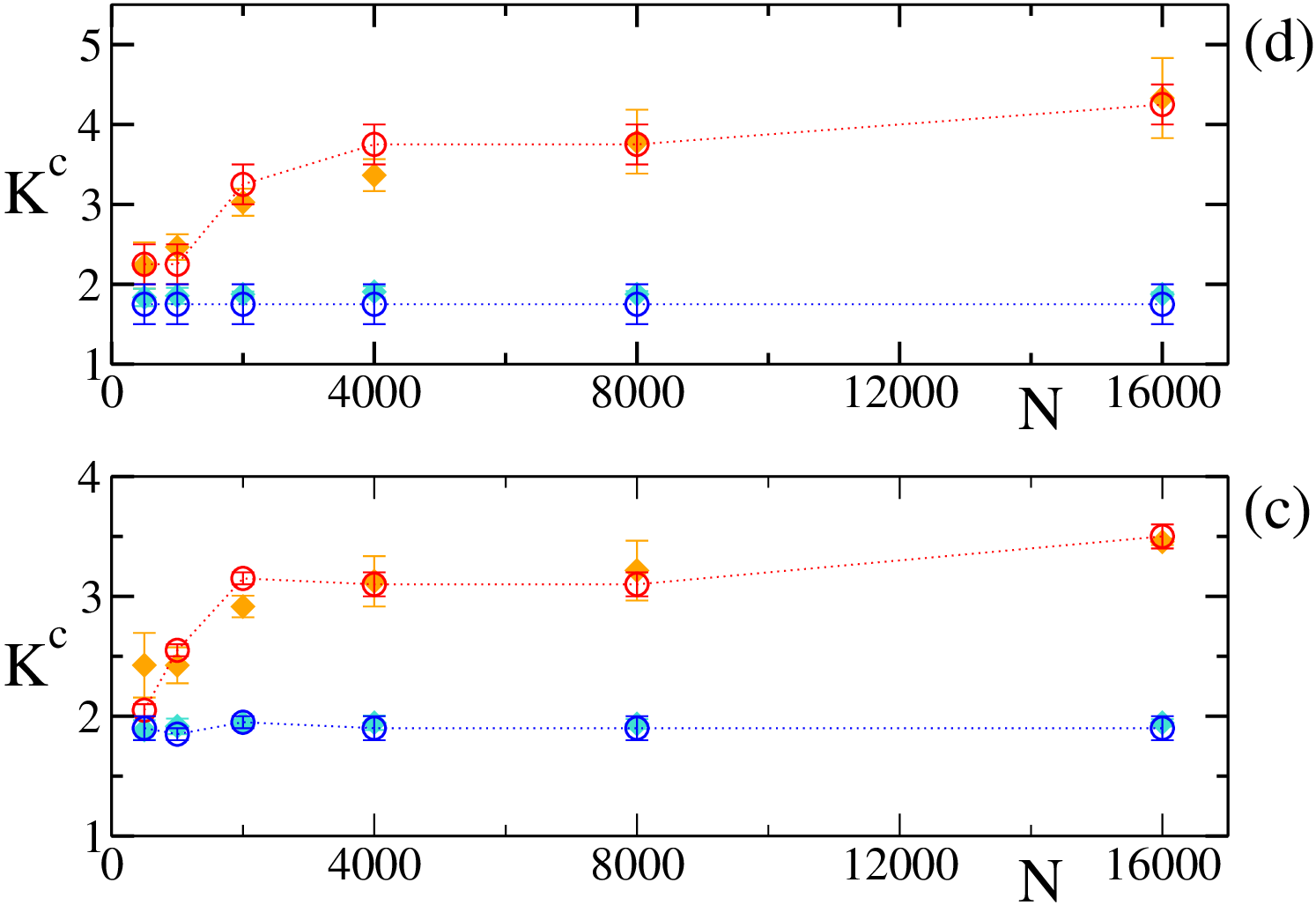}
\caption{(Color Online) \red{Critical couplings $K_1^c$ (orange filled diamonds and red empty circles) 
and $K_2^c$ (cyan filled diamonds and blue empty circles) versus the system size $N$: 
(a) $m=0.8$, (b) $m=1$, (c) $m=2$, and (d) $m=6$. 
The filled symbols refer to estimates performed with protocol (S), while
empty symbols, in panels (c) and (d), have been obtained with protocol (I) (protocol (II)) for $K^c_1$ ($K_2^c$).
The dashed (black) lines in panels (a) and (b) are the mean field 
values $K_1^{MF}$. This quantity is not reported in panels (c) and (d) 
for clarity reasons, due to its large value, namely, $K_1^{MF}=5.31$ for $m=2$ and $K_1^{MF}=13.27$ for $m=6$.
For all panels the data have been derived by averaging  in time over a window $T_W=2,000$ and over 8 (5) 
different initial conditions for the protocol (S) (protocol (I) and (II)). For each simulation an initial
a transient time $T_R \simeq 20,000$ ($T_R \simeq 1,000 - 80,000$) has been discarded for protocol 
(S) (protocol (I) and (II)).}
\label{Kc}
}
\end{figure}

\begin{figure}
\includegraphics[angle=0,width=8.cm]{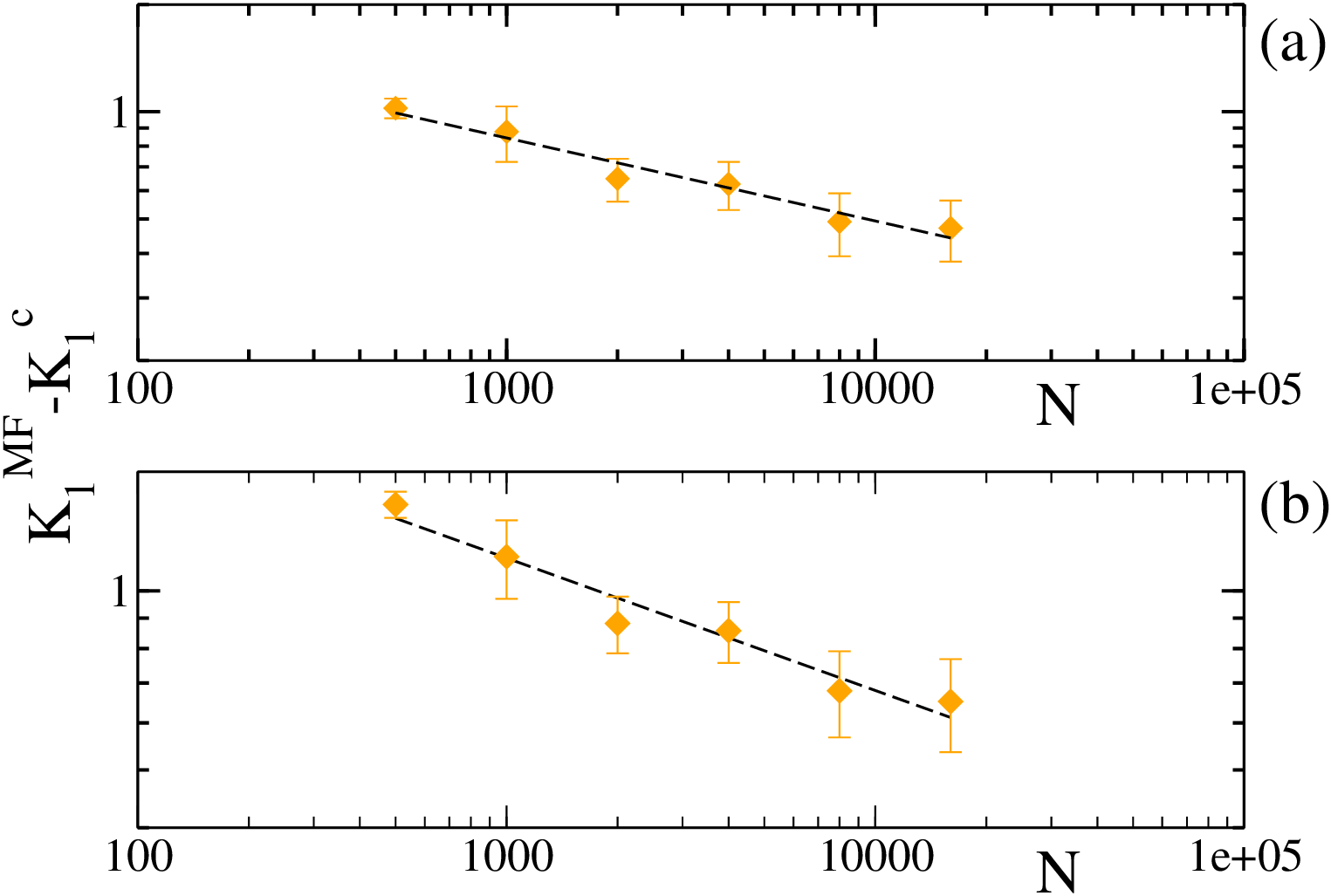}
\caption{(Color Online) \red{The differences $K_1^{MF}-K_1^c(N)$  (filled orange diamonds) are reported versus 
the system size $N$ for (a) $m=0.8$ and (b) $m=1$.  
The dashed (black) lines in both panels are power-law fits to the data: the difference vanishes
as $0.42 \times N^{-0.23}$ for $m=0.8$ (a) and as $5.27 \times N^{-0.22}$. for $m=1$ (b).
The data for $K_1^c$ are the same reported in panel (a) and (b) in Fig. \ref{Kc}.}
\label{KcFit}
}
\end{figure}

\red{The results are reported in Fig.~\ref{Kc} for four different values of the mass;
it is clear, by looking at the data displayed in Figs.~\ref{Kc}(c) and (d), 
that protocol (I) (protocol (II)) and protocol (S)  give essentially the same critical 
couplings, suggesting that their values are not dependent on the chosen initial conditions. 
Furthermore,  while $K_2^c$ reveals a weak dependence on $N$, $K_1^c$ increases steadily with the system size. 
On the basis of our numerical data, it seems that the growth slow down at large $N$, 
but we are unable to judge if $K_1^c$ is already saturated to an asymptotic 
value at the maximal reached system size, namely $N=16,000$.}
\red{ 
To clarify this issue we compare our numerical results for $K_1^c$ with the mean field estimated 
$K_1^{MF}$ reported in Eq. (\ref{K1c}). The mean field result is always larger than the
finite size measurements, however for small masses, namely $m=0.8$ and $m=1.0$,
$K_1^c$ seems to approach this asymptotic value already for the considered number of oscillators,
as shown in Figs.~\ref{Kc}(a) and (b). Therefore, in these two cases 
we attempt to identify the scaling law ruling the approach of  $K_1^c$ to
its mean field value for increasing system sizes. 
The results reported in Fig.~\ref{KcFit} suggest the following power law
\begin{equation}
\label{k1finiteN}
[K_1^{MF}-K_1^c(N)] \propto N^{-\gamma}
\quad ;
\end{equation}
with $\gamma \simeq 0.22 - 0.23$.
}

\begin{figure}
\includegraphics[angle=0,width=8.cm]{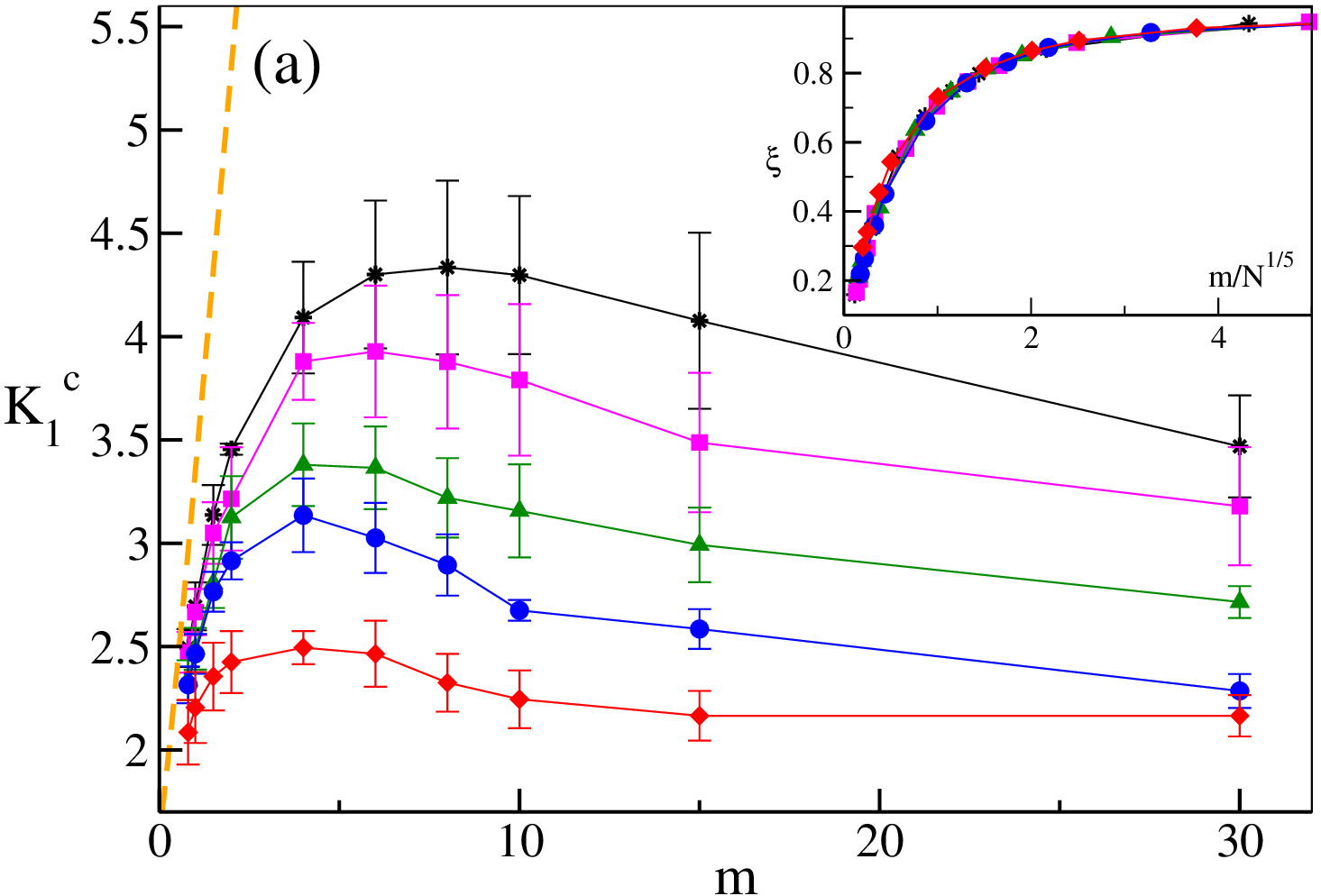}
\includegraphics[angle=0,width=8.cm]{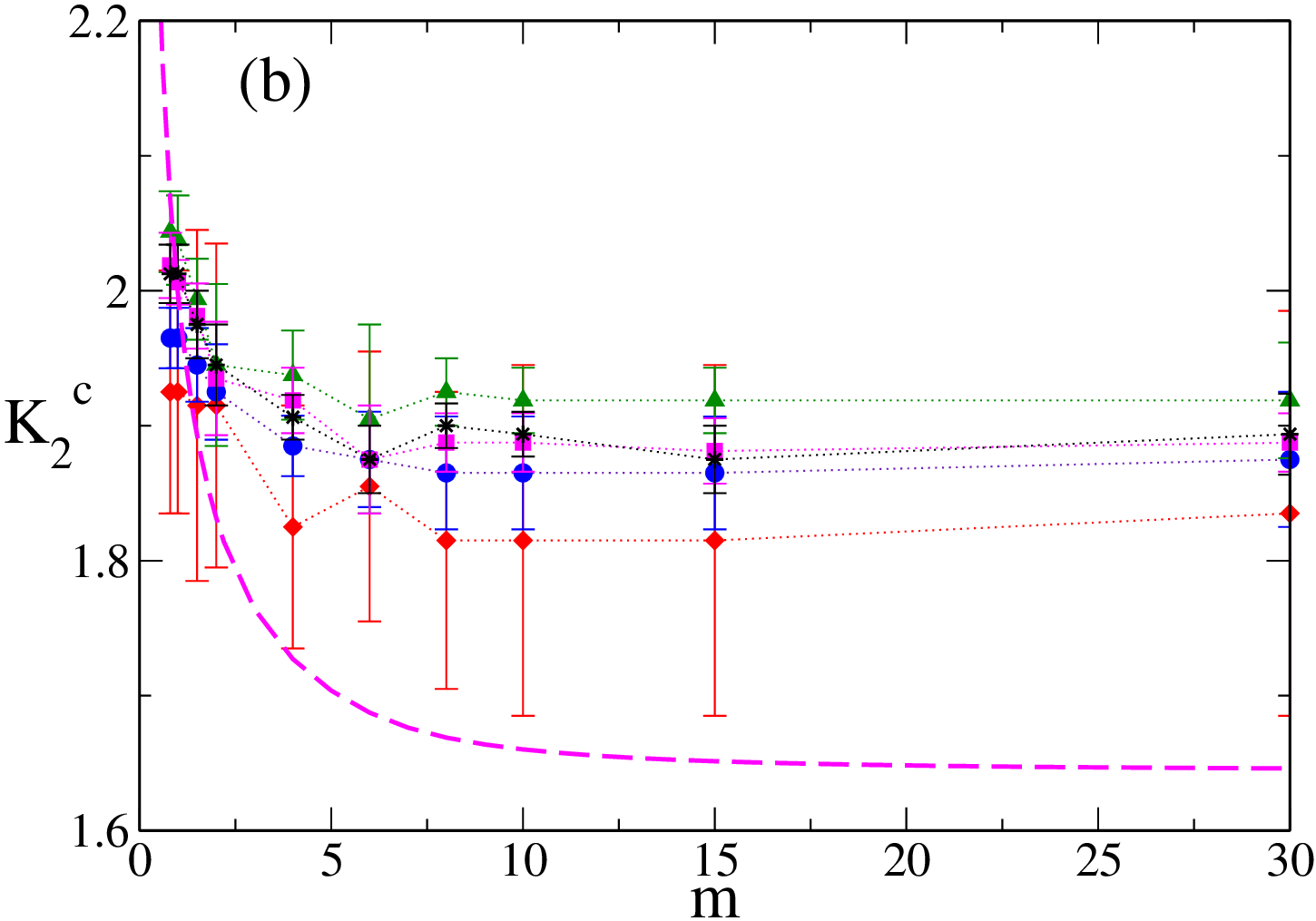}
\caption{(Color Online) Critical couplings $K_1^c$ (a) and $K_2^c$ (b) versus the mass $m$
for different system sizes $N$. Namely, $N=1,000$ (red diamond), 2,000 (blue circles),
4,000 (green triangles), 8,000 (magenta squares) and 16,000 (black asteriskes).
\red{The dashed (orange) line in (a) is the mean field  estimates $K_1^{MF}$;
while the dashed (magenta) line in (b) is the value $K^{II}_{m}$ obtained 
by the TLO theory. The inset in panel (a) report the critical rescaled couplings 
$\xi =(K_1^{MF} -K_1^c)/K_1^{MF}$ as a function of $m/N^{1/5}$.}
The estimates have been obtained with protocol (S),  by averaging 
in time over a window $T_W=2,000 - 4,000$ and over 8 different initial conditions. 
For each simulation  an initial transient time $T_R \simeq 20,000$ has been discarded. 
\label{Kcmass}
}
\end{figure}

Let us now consider several different values of the mass in the range $0.8 \le m \le 30$; the data
for the critical couplings are reported in Fig~\ref{Kcmass} for different system sizes
ranging from $N=1,000$ to $N=16,000$. It is evident that $K_1^c$ grows with $N$ for all masses,
\red{while $K_2^c$  varies in a more limited manner.  In particular,
the estimated $K_2^c$ shows an initial decrease with $m$ followed by a constant plateau at
larger masses (as shown in Fig~\ref{Kcmass} (b)).
A possible mean field estimation for $K_2^c$ can be given by 
the minimal value $K^{II}_{m}$ reached by the coupling along the TLO curve $r^{II}(K)$.
This value is reported in Fig~\ref{Kcmass} (b) together with the finite size data:
at small masses $K^{II}_{m}$ gives a reasonable approximation of the numerical
data, while at larger masses it is always smaller than the finite size results
and it saturates to a constant value for $m \to \infty$.
These results indicate that finite size fluctuations destabilizes the coherent state
at larger coupling than those expected from a mean field theory.  
}
 
\red{
On the other hand $K_1^c$ appears to increase with $m$ up to some maximal
value and then to decrease at large masses. 
However, this is clearly a finite size effect,
since by increasing $N$ the position of the maximum shifts to larger masses.
The finite size curves $K_1^c=K_1^c(m,N)$
are always smaller than the mean field result $K_1^{MF}$ (dashed orange line in Fig~\ref{Kcmass} (a))
for all considered system sizes and masses. However, as shown in the inset of 
Fig~\ref{Kcmass} (a), such curves collapse one over the other if the variables
are properly rescaled, suggesting the following  functional dependence}
\red{
\begin{equation}
\label{k1scale}
\xi \equiv \frac{K_1^{MF}-K_1^c(m,N)}{K_1^{MF}} = G\left(\frac{m}{N^{\gamma}}\right)
\quad ;
\end{equation}
where $\gamma = 1/5$. This result is consistent with the values of the scaling exponent 
$\gamma$ found for fixed mass by fitting the data with the expression reported in Eq.~(\ref{k1finiteN}).
However, we are unable to provide any argument to justify such scaling and
further analysis are required to intepret these results. A possible strategy could
be to extend the approach reported in ~\cite{hong2007} for the finite size analysis of the
usual Kuramoto transition to the Kuramoto model with inertia.
}

\subsection{Drifting Clusters}

\begin{figure}
\includegraphics[angle=0,width=8.cm]{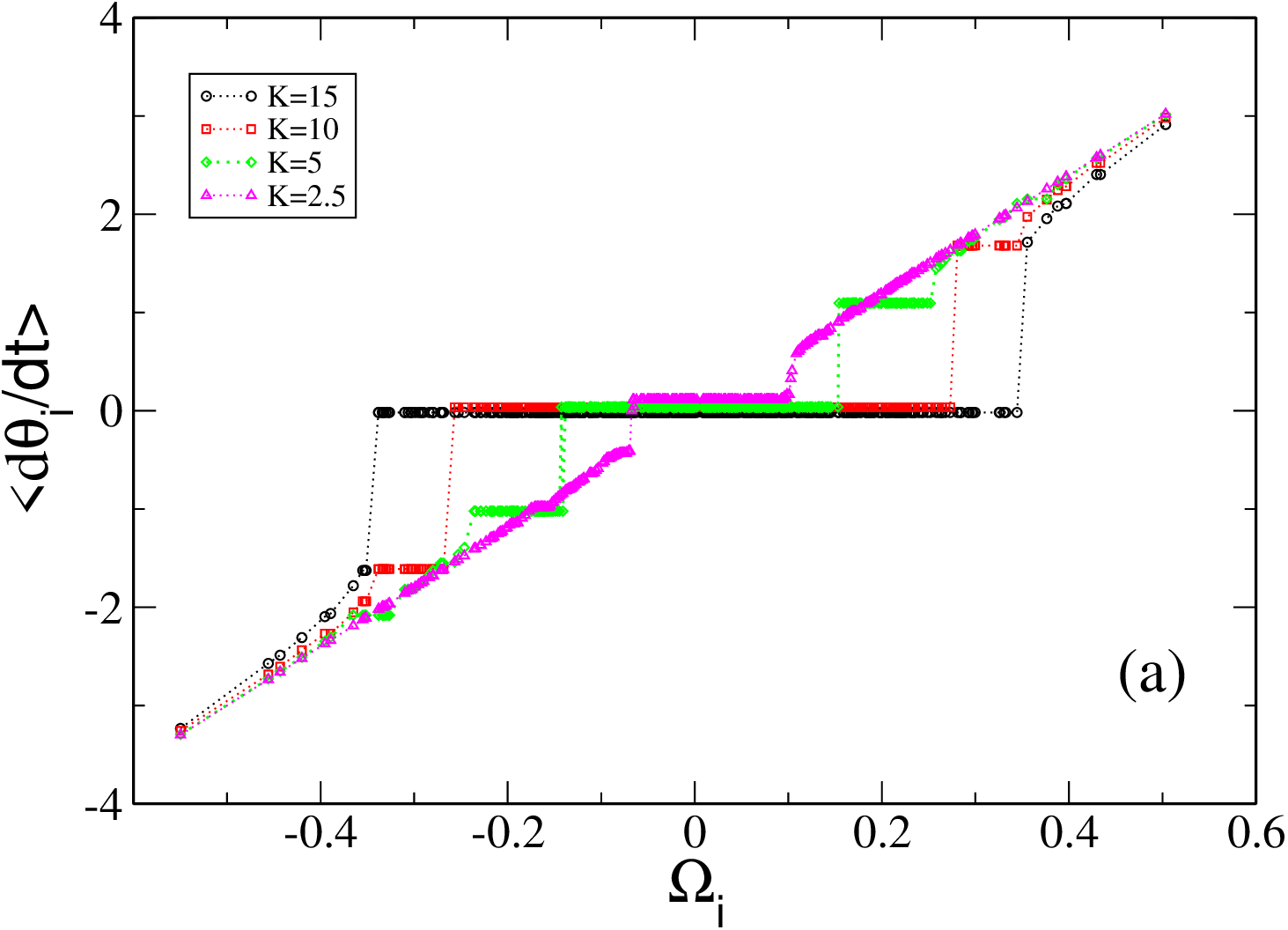}
\includegraphics[angle=0,width=8.cm]{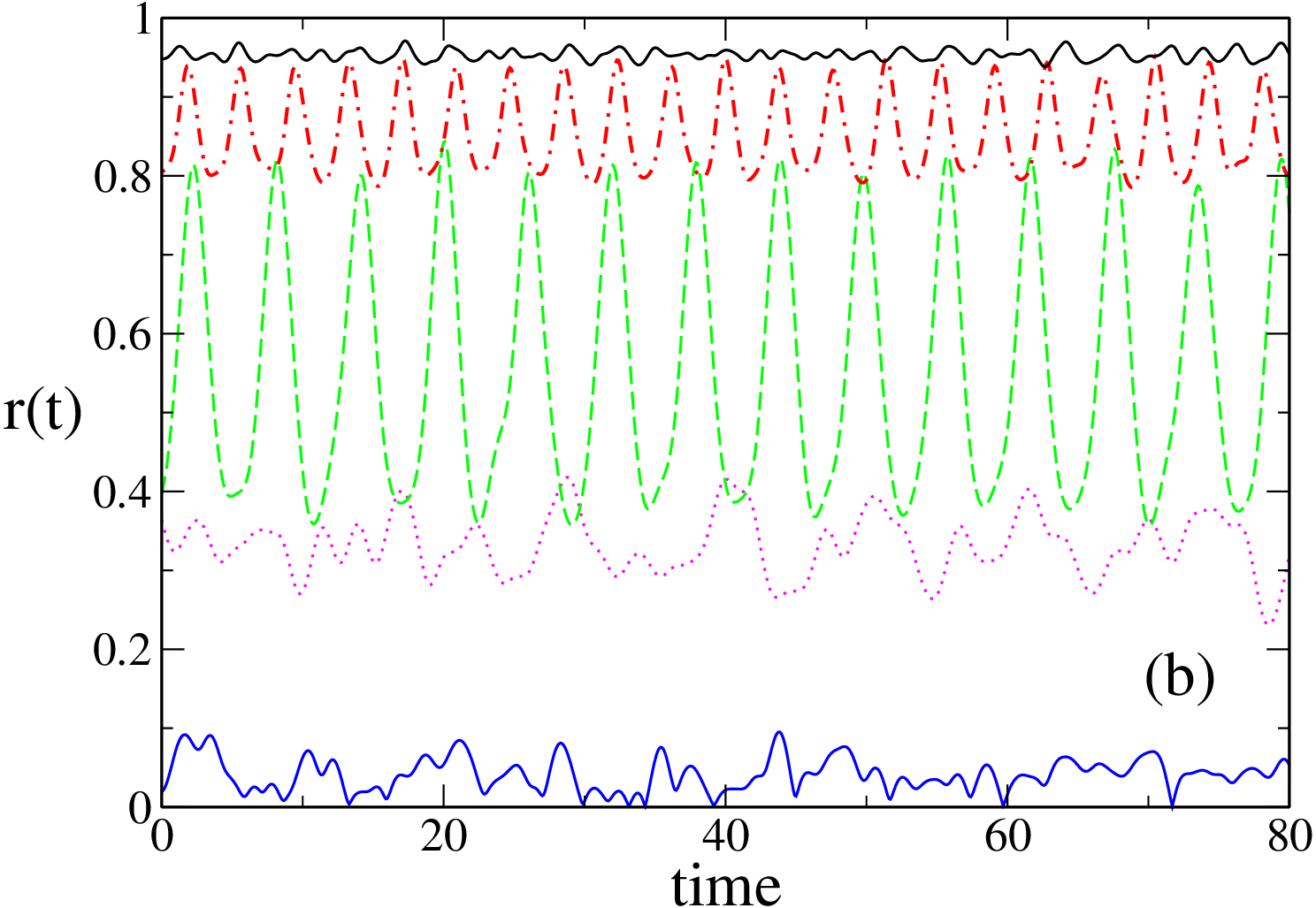}
\caption{(Color Online) 
(a) Average phase velocity $\langle \dot \theta_i \rangle$ of the oscillators 
versus their natural frequencies $\Omega_i$: (magenta) triangles refer to $K=2.5$, (green) diamond 
to $K=5$, (red) squares to $K=10$ and (black) circles to $K=15$. For each simulation  an initial transient $T_R \simeq 5,500$ 
has been discarded and the estimates have been obtained with protocol (I), by averaging in time over a window $T_W=5,000$. 
(b) Order parameter $r(t)$ versus time for $m=6$ and $N=500$ and different coupling constants $K$:
the (blue) solid curve corresponds to $K=1$; the (magenta) dot-dashed line to $K=2.5$,
the (green) dashed line to $K=5$, the (red) dashed line to $K=10$ and the (black) solid line
to $K=15$. The data have been obtained by employing protocol (I) and 
for each simulation  an initial transient time $T_R \simeq 1,500$ has been discarded
\red{and data are averaged over a time $T_W=5000$}.
\label{Plateau}
}
\end{figure}

 As already noticed in \cite{tanaka1997first}, for sufficiently 
large value of the mass one observes that the partially synchronized phase, 
obtained by following protocol (I), is characterized not only by the presence
of the cluster of locked oscillators with $\langle \dot \theta \rangle \simeq 0$, 
but also by the emergence of clusters composed by drifting oscillators
with finite average velocities. This is particularly clear in Fig.~\ref{Plateau} (a),
where we report the data for mass $m=6$. By increasing the coupling $K$
one observes for $K > 3$ the emergence of a cluster of whirling oscillators with
a finite velocity $|\langle \dot \theta \rangle| \simeq 1.05$, these oscillators
have natural frequencies in the range $|\Omega_i| \simeq 0.15-0.25$. The number of oscillators 
in this secondary cluster $N_{DC}$ increases up to $K \simeq 5$, then it declines, 
finally the cluster is absorbed in the main locked group for $K \simeq 7$.  At the same time a
second smaller cluster emerges characterized by a larger average velocity $|\langle \dot \theta \rangle| \simeq 1.6$
(corresponding to larger $|\Omega_i| \simeq 0.27-0.34$). 
This second cluster merges with the locked oscillators 
for $K \simeq 12.5$, while a third one, composed of oscillators
with even larger frequencies $|\Omega_i|$ and characterized by larger average
phase velocity, arises. This process repeats until the full synchronization of the
system is achieved.

\begin{figure}
\includegraphics[angle=0,width=8.cm]{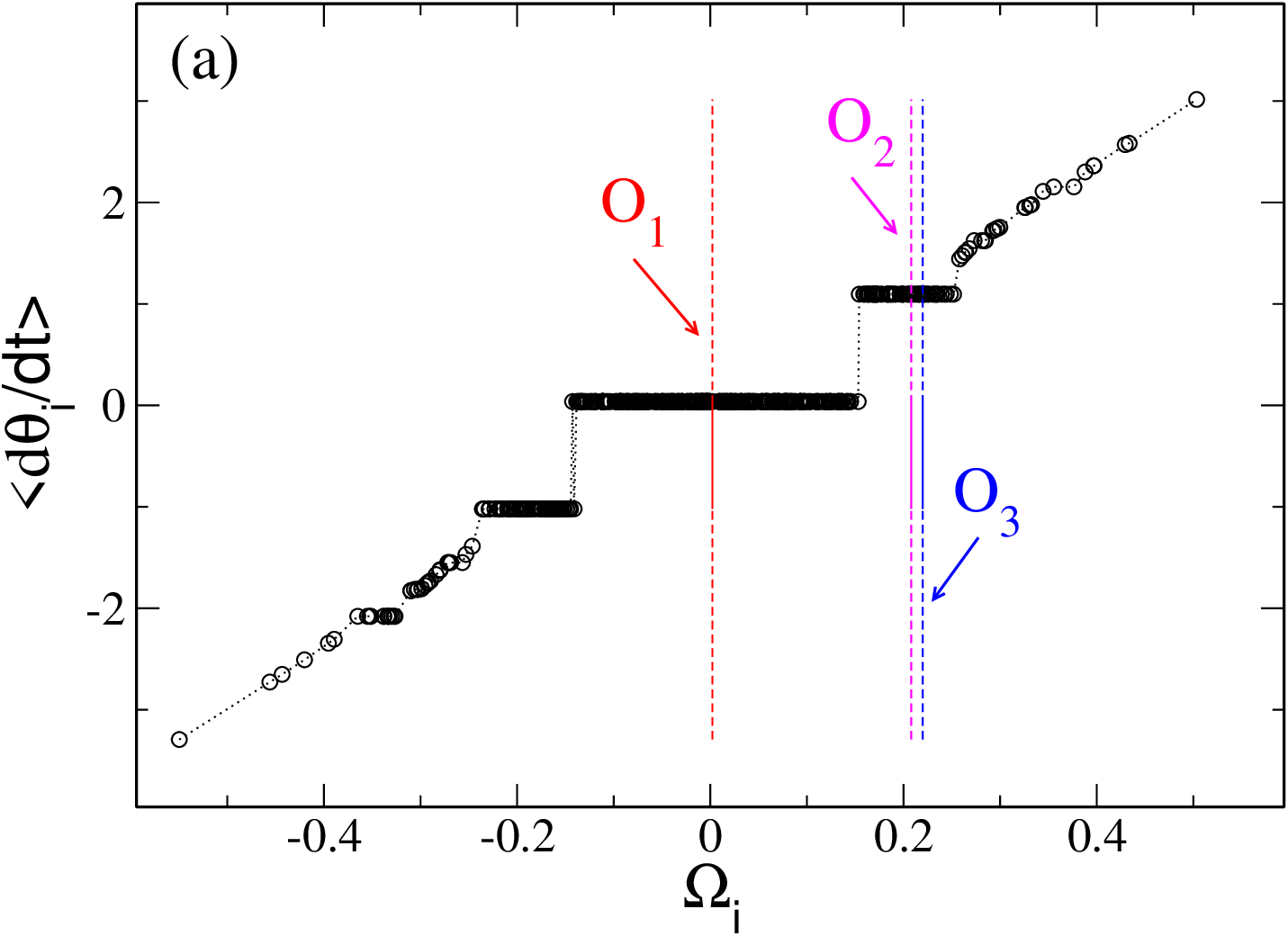}
\includegraphics[angle=0,width=8.cm]{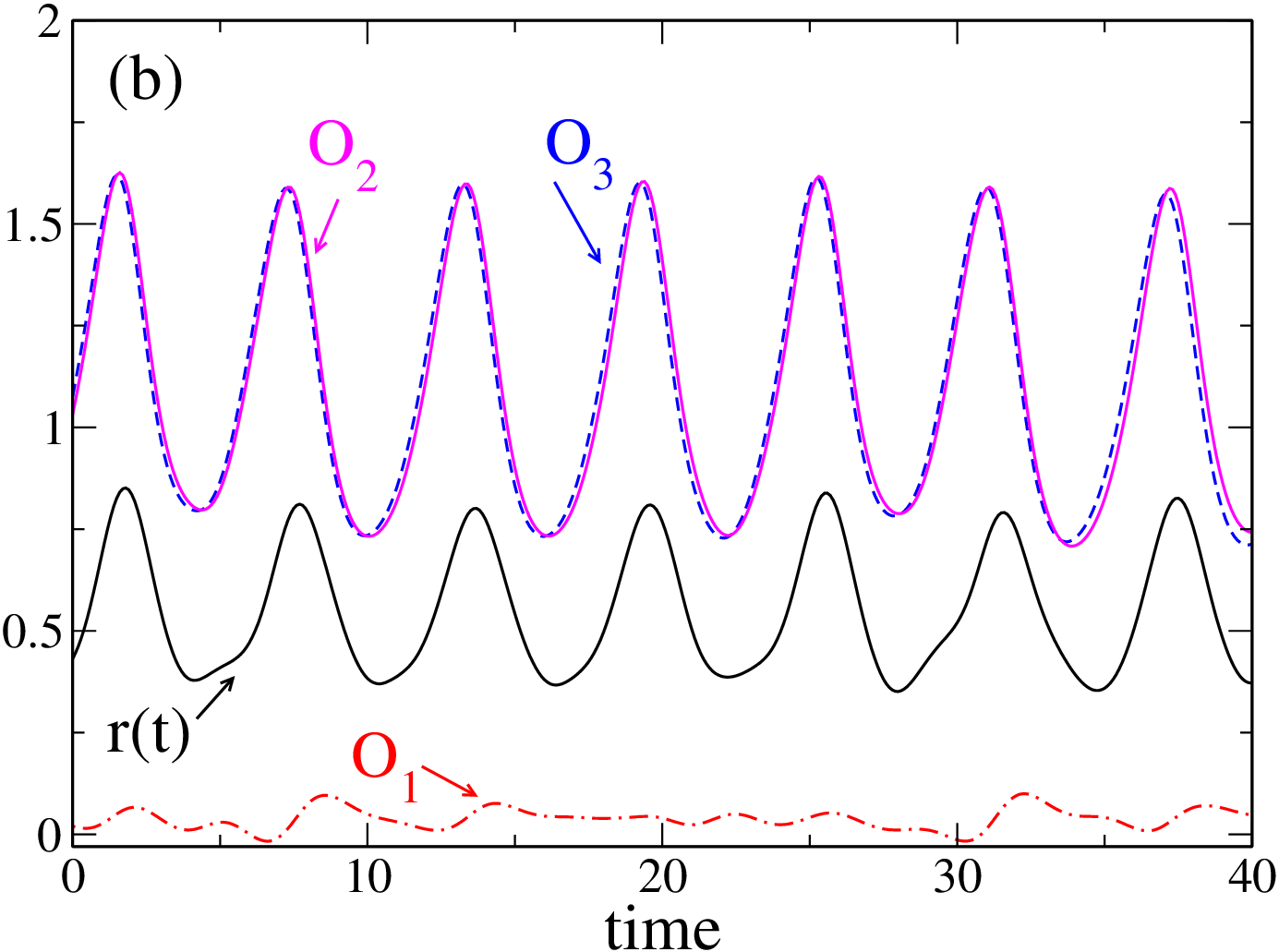}
\caption{(Color Online) 
(a) Average phase velocity $\langle \dot \theta_i \rangle$ of the oscillators
versus the corresponding natural frequency $\Omega_i$.
The vertical dashed lines denote the three oscillators,  $O_1$, $O_2$ and $O_3$,
whose dynamical evolution is shown in (b).
(b) The black curve represents the order parameter ${r}$ versus time,
the other curves refer to the time evolution of the phase velocities $\dot \theta(t)$ 
of the three oscillators $O_1$ (red dot-dashed curve), $O_2$ (magenta solid line) 
and $O_3$ (dashed blue curve). 
For each simulation  an initial transient time $T_R \simeq 1,000$ has been discarded,
the averages reported in (a) have been obtained over a time window $T_W=20,000$.
In both panels $K=5$, $m=6$ and $N=500$.  
\label{K5}
}
\end{figure}

The effect of these extra clusters on the collective dynamics is to induce 
oscillations in the temporal evolution of the order parameter, as one can see from Fig.~ \ref{Plateau} (b). 
In presence of drifting clusters characterized by the same
average velocity (in absolute value), as for $m=6$ and $K=5$ in Fig.~ \ref{Plateau} (b), 
$r$ exhibits almost regular oscillations and the period of these oscillations is related to the 
one associated to the oscillators in the whirling cluster. This can be appreciated from Fig.~ \ref{K5} (b), where 
we compare the evolution of the istantaneuous velocity $\dot \theta_i$ for three oscillators and the time course of $r(t)$.
We consider one oscillator $O_1$ in the locked cluster,
and 2 oscillator $O_2$ and $O_3$ in the drifting cluster.
We observe that these latter oscillators display essentially synchronized motions,
while the phase velocity of $O_1$ oscillates irregularly around zero. Furthermore,
the almost periodic oscillations of the order parameter $r(t)$ are clearly
driven by the periodic oscillations of $O_2$ and $O_3$ (see Fig.~ \ref{K5} (b)). 

\begin{figure}
\includegraphics[angle=0,width=8.cm]{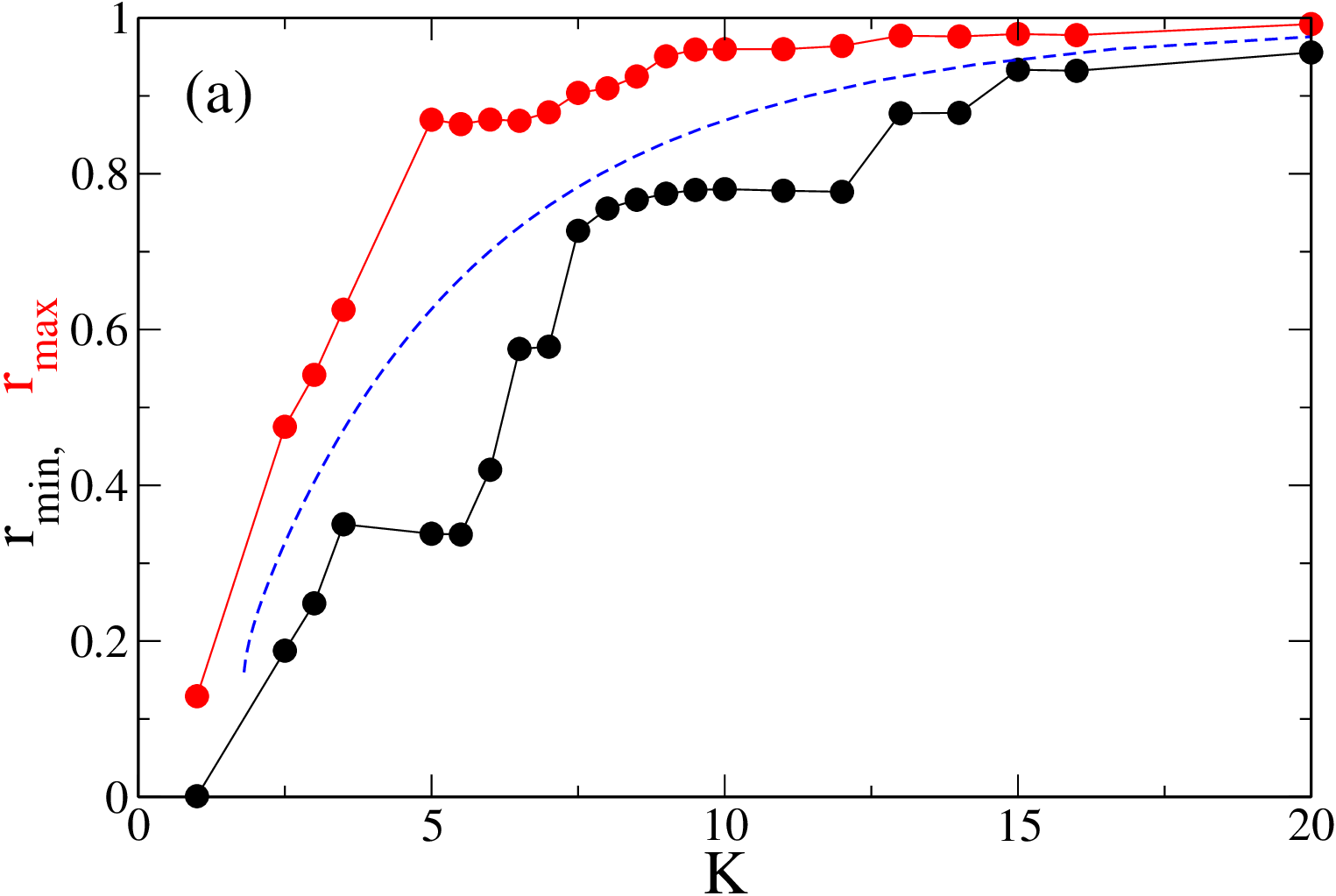}
\includegraphics[angle=0,width=7.5cm]{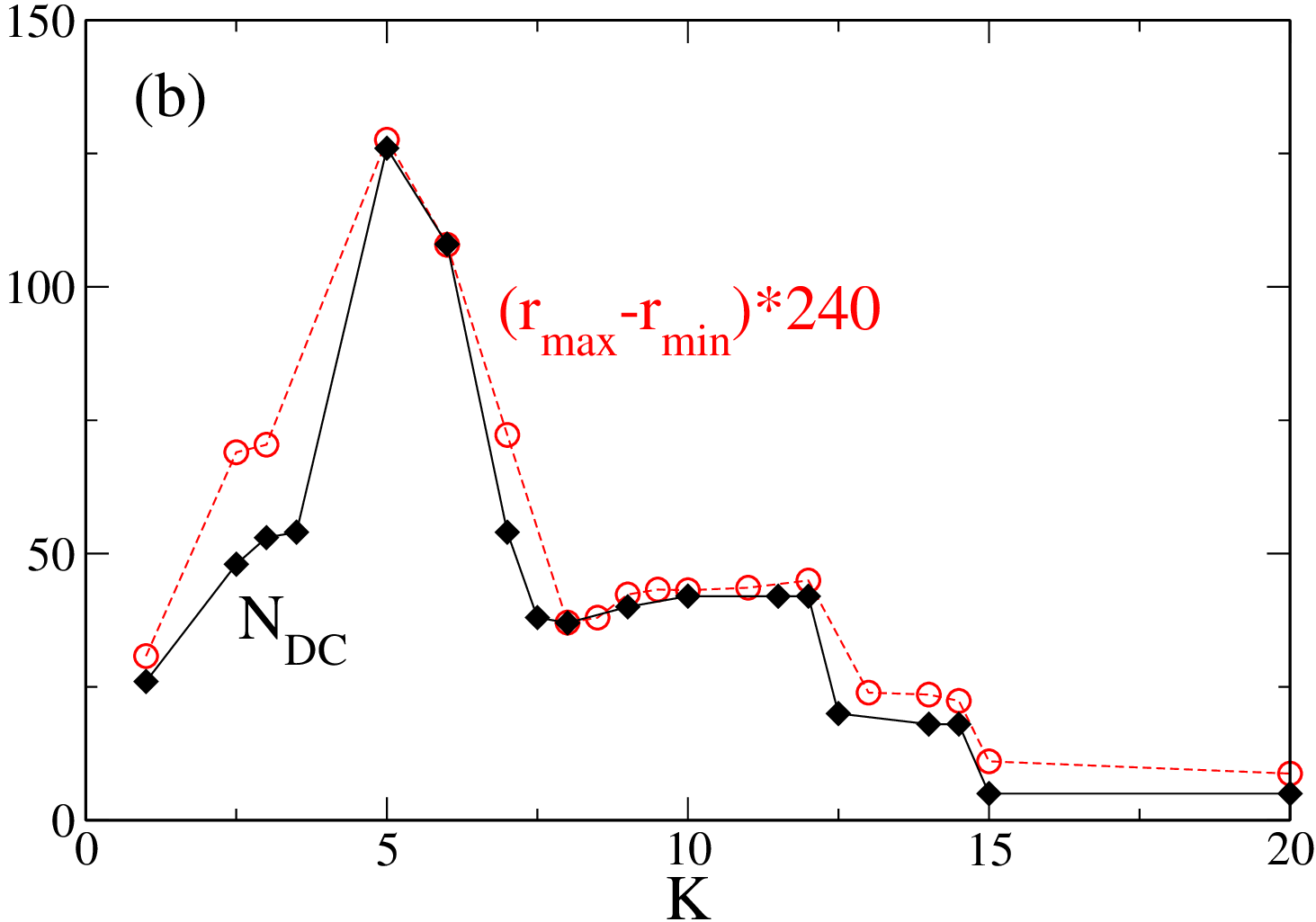}
\caption{(Color Online) 
(a) Minima and maxima of the order parameter $r$ as a function of the 
coupling constant $K$. The (blue) dashed line refer to the theoretical
estimate $r^I$, as obtained by employing Eqs. (\ref{rL}) and (\ref{rD}).
(b) The number of oscillators in the drifting clusters $N_{DC}$
(filled black diamond) is reported versus the coupling $K$ together
with the amplitude of the oscillations of the order parameter
$r_{max}-r_{min}$ (empty red circles) rescaled by a factor 240.
For each simulation  an initial transient time $T_R \simeq 1,500$ has been discarded.
The estimates have been obtained with protocol (I),  by averaging 
in time over a window $T_W=5000$, $m=6$, $N=500$.
\label{RminRmax}
}
\end{figure}

We have also verified that the amplitude of the oscillations of $r(t)$ (measured
as the difference between the maximal $r_{max}$ and the minimal $r_{min}$ value
of the order parameter) and the number of oscillators in the drifting clusters
$N_{DC}$ correlates in an almost linear manner, as shown in \red{Fig.~\ref{RminRmax} (b)}.
Therefore we can conclude that the oscillations 
observable in the order parameter are induced by the presence of large  secondary clusters 
characterized by finite whirling velocities. At smaller masses (e.g. $m=2$)
oscillations in the order parameter are present, but they are much more smaller and irregular
(data not shown). These oscillations are probably due to finite size effects, since in this case
we do not observe any cluster of drifting oscillators in the whole range from 
asynchronous to fully synchronized state.

The situation was quite different in the study reported in 
\cite{tanaka1997self}, where the authors considered natural frequencies
$\{\Omega_i\}$ uniformly distributed over a finite interval and not Gaussian 
distributed as in the present study. In that case, by considering an initially clusterized state, 
similar to what done for protocol (S), $r(t)$ revealed regular oscillations even for masses as small as $m=0.85$.
In agreement with our results, the amplitude of the oscillations measured in \cite{tanaka1997self} 
decreases by approaching the fully synchronized state (as shown in Fig.\ref{RminRmax}). However, 
the authors \red{in~\cite{tanaka1997self}} did not relate the observed oscillations 
in $r(t)$ with the formation of drifting clusters.

As a final aspect, as one can appreciate from Fig.~\ref{finiteN}, for larger 
masses the discrepancies between the measured $\bar r$, obtained by employing
protocol (I), and the theoretical mean field result $r^I$ increase. 
In order to better investigate the origin of these discrepancies,
we report in Fig.\ref{RminRmax} the minimal
and maximal value of $r$ as a function of the coupling $K$ and we compare these
values to the estimated mean field value $r^I$.  The comparison clearly reveals
that $r^I$ is always contained between $r_{min}$ and $r_{max}$, therefore 
the mean field theory captures correctly the average increase of the order parameter,
but it is unable to foresee the oscillations in $r$. A new version of the theory developed by TLO
in \cite{tanaka1997first} is required in order to include also the effect of clusters of whirling
oscillators. \red{A similar synchronization scenario, where oscillations in $r(t)$ are
induced by the coexistence of several drifting clusters, has been recently reported 
for the Kuramoto model with degree assortativity~\cite{restrepo}.}

\section{Diluted networks}
\label{sec5}

In this Section we will analyze diluted neural networks obtained by
considering random realizations of the coupling matrix $C_{i,j}$ with
the constraints that the matrix should remain symmetric \red{and the
in-degree should be constant and equal to $N_c$} \footnote{\red{In particular,
each row $i$ of the coupling matrix $C_{i,j}$ is generated by choosing randomly a node $m$
and by imposing $C_{i,m}= C_{m,i} = 1$; this procedure is repeated 
until $N_c$ elements of the row are set equal to one. Obviously,
before accepting a new link, one should verify that in the considered
row the number of links is smaller than $N_c$ and that this is true
also for all the interested columns. Finally, we have performed an iterative 
procedure to ensure that all rows and columns contain 
exactly $N_c$ non zero elements.}}.
In particular, we will examine if the introduction of the random dilution
in the network will alter the results obtained by the mean-field theory
and if the transition will remains hysteretic or not. For this
analysis we limit ourselves to a single value of the mass, namely $m=2$.

\begin{figure}
\includegraphics[angle=0,width=8.cm]{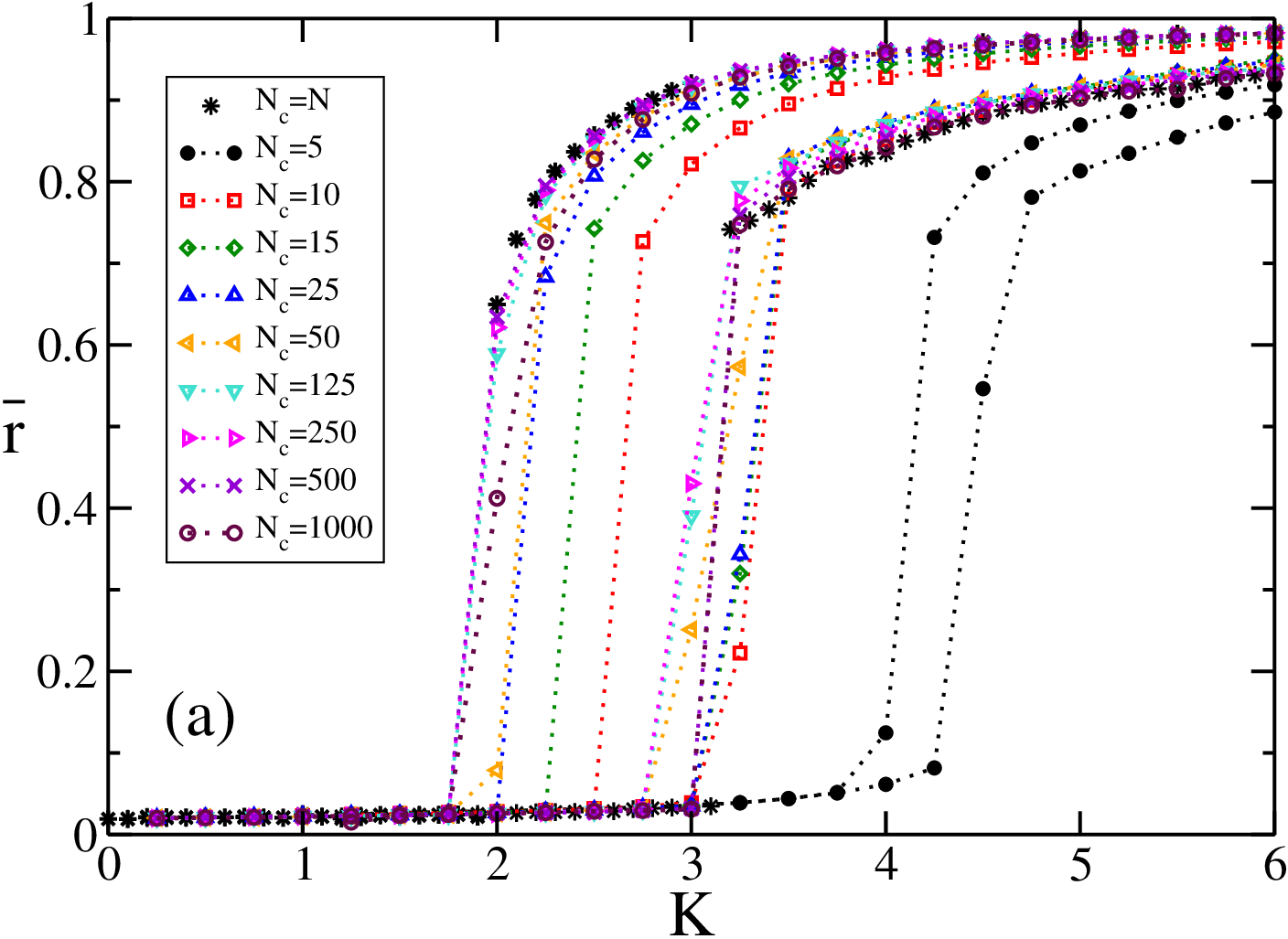}
\includegraphics[angle=0,width=8.cm]{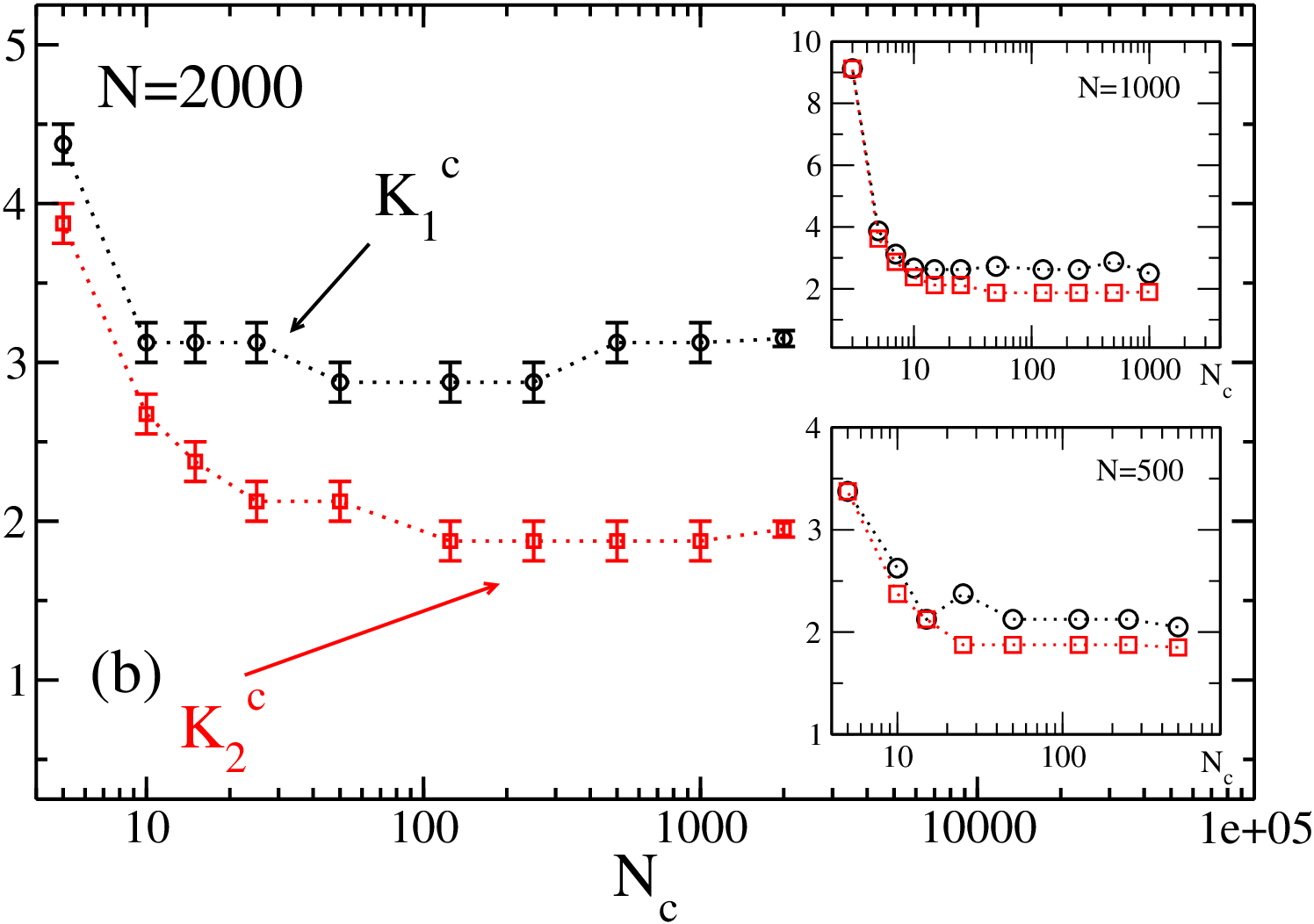}
\caption{(Color Online) (a) Average order parameter ${\bar r}$ versus the coupling constant $K$
for diluted neural networks for various $N_c$: $5$ (filled black circlesa); 10 (red squares);
15 (green diamond); 25 (blue up triangles); 50 (orange left triangles); 125 (turquoise down triangles);
250 (right magenta triangles); 500 (violet crosses); 1,000 (empty maroon circles); 2,000 (black asteriskes).
(b) Critical constants $K_1^c$ and $K_2^c$ estimated for various values of the in-degree.
The numerical data refer to $N=2,000$; the upper inset refer to $N=1,000$, the lower one to $N=500$.
For all simulations $m=2$, $T_R=10,000$, and $T_W=2,000$;  each series of simulations 
have been obtained by following protocol (I) and then (II) starting from $K=0$ until $K_M =20$ 
with steps $\Delta K=0.25$. The reported data have been obtained by averaging 
over 10 - 20 different series of simulations, each corresponding to a different realization 
of the random network and of the distribution of the frequencies $\{\Omega_i\}$.  
The error bars in panel (b) correspond to $\Delta K/2$.
\label{diluita2000}}
\end{figure}

\begin{figure}
 \includegraphics[angle=0,width=8.cm]{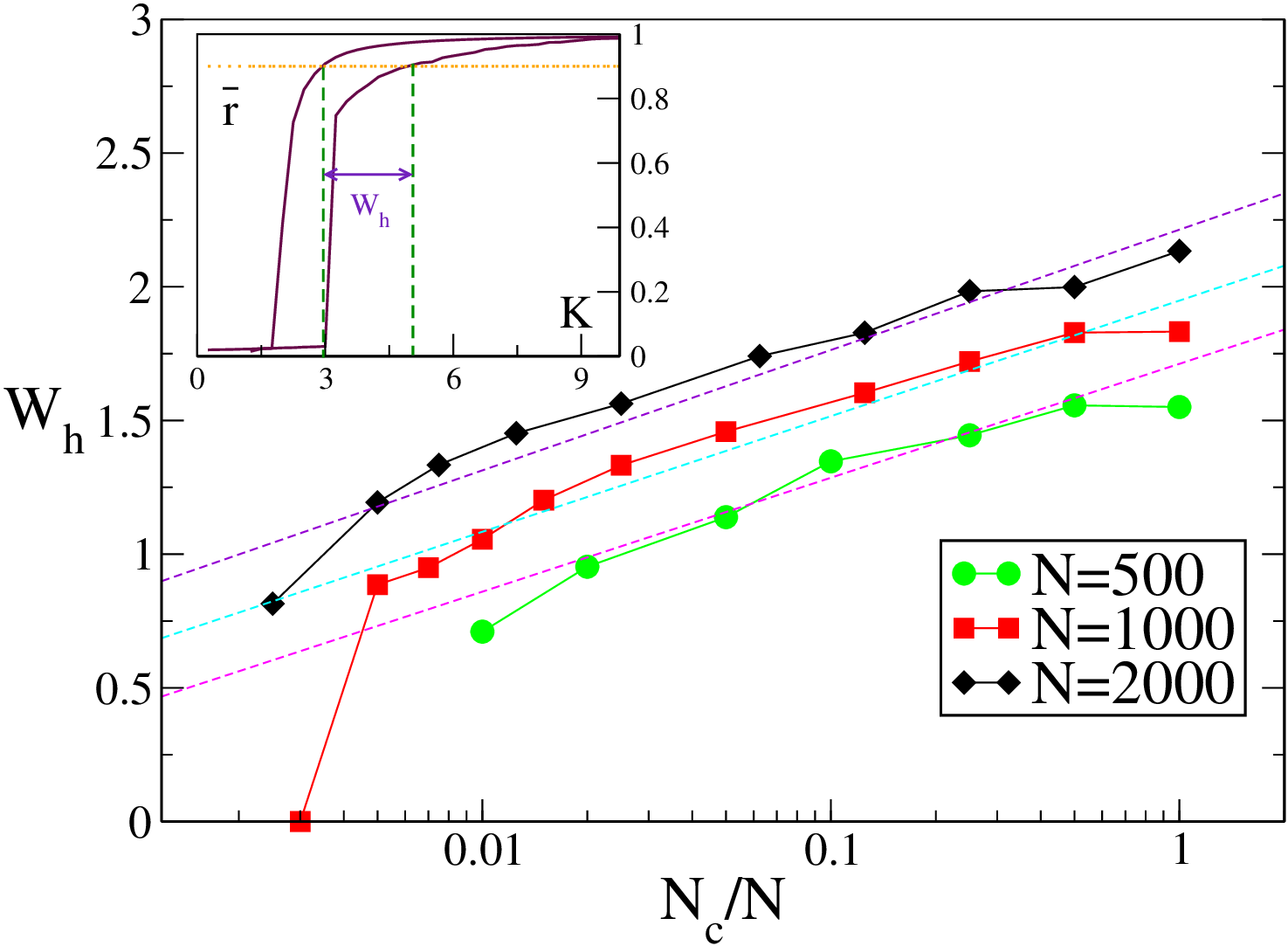}
 \caption{(Color Online) Width of the hysteretic loop $W_h$, measured in correspondence
of a order parameter value $\bar{r}=0.9$, as a function of the percentage of
connected links $N_c/N$. The (green) circles refer to $N=500$,
the (red) squares to $N=1000$ and the (black) diamond to $N=2,000$.
The dashed lines refer to logarithmic fitting to the data
in the range $0.01 < N_c/N \le 1$. 
\red{In the inset is graphically explained how $W_h$ has been estimated, 
starting from one of the curves reported in Fig.~\ref{diluita2000} (a).} The data refer 
to the same parameters and simulation 
protocols as in Fig.~\ref{diluita2000}.
 \label{ampiezzaisteresi}}
\end{figure}

Let us first consider how the dependence of the order parameter $\bar r$
on the coupling constant $K$ will be modified in the diluted systems.
In particular, we examine the outcomes of simulations performed 
with protocol (I) and (II) for a system size $N=2,000$ and different realizations
of the diluted network ranging from the fully coupled case to $N_c=5$.
The results, reported in Fig.~\ref{diluita2000}, reveal that as far $N \ge 125$
(corresponding to the $\simeq 94 \%$ of cutted links) it is difficult to distinguish
among the fully coupled situation and the diluted ones. The small observed discrepancies 
can be due to finite size fluctuations. For larger dilution, the curves obtained with protocol (II)
reveal a more rapid decay at larger coupling. Therefore $K_2^c$ increases by decreasing $N_c$
and approaches $K_1^c$ as shown in Fig.~\ref{diluita2000} (b). The dilution has almost no
effect on the curve obtained with protocol (I), in particular 
$K_1^c$ remains unchanged (apart fluctuations within the error bars) 
until \red{the percentage of incoming links $N_c/N$ reduces below the $0.5 \%$}. For smaller
connectivities both $K_1^c$ and $K_2^c$ shift to larger coupling and they approach
one another, indicating that the synchronization transition from hysteretic tends 
to become continuous. Indeed this happens for $N=1,000$ and $N=500$ 
(as shown in the inset of Fig.~\ref{diluita2000} (b)): 
for such system sizes we observe essentially the same scenario as for $N=2,000$, 
but already for in-degrees $N_c \le 5$ the transition is no more hysteretic. This seems to suggest
that by increasing the system size the transition will stay hysteretic for vanishingly
small percentages of connected  (incoming) links. \red{This is confirmed by
the data shown in Fig.~\ref{ampiezzaisteresi}, where we report the width
of the hysteretic loop $W_h$, measured at a fixed value of the order parameter, 
namely we considered $\bar r=0.9$. For increasing system sizes $W_h$,
measured for the same fraction of connected links $N_c/N$, increases, while  
the continuous transition, corresponding to $W_h \equiv 0$, 
is eventually reached for smaller and smaller value of $N_c/N$.}
Unfortunately, due to the CPU costs, we are unable to investigate in details 
diluted systems larger than $N=2,000$.

\begin{figure}
\includegraphics[angle=0,width=8.cm]{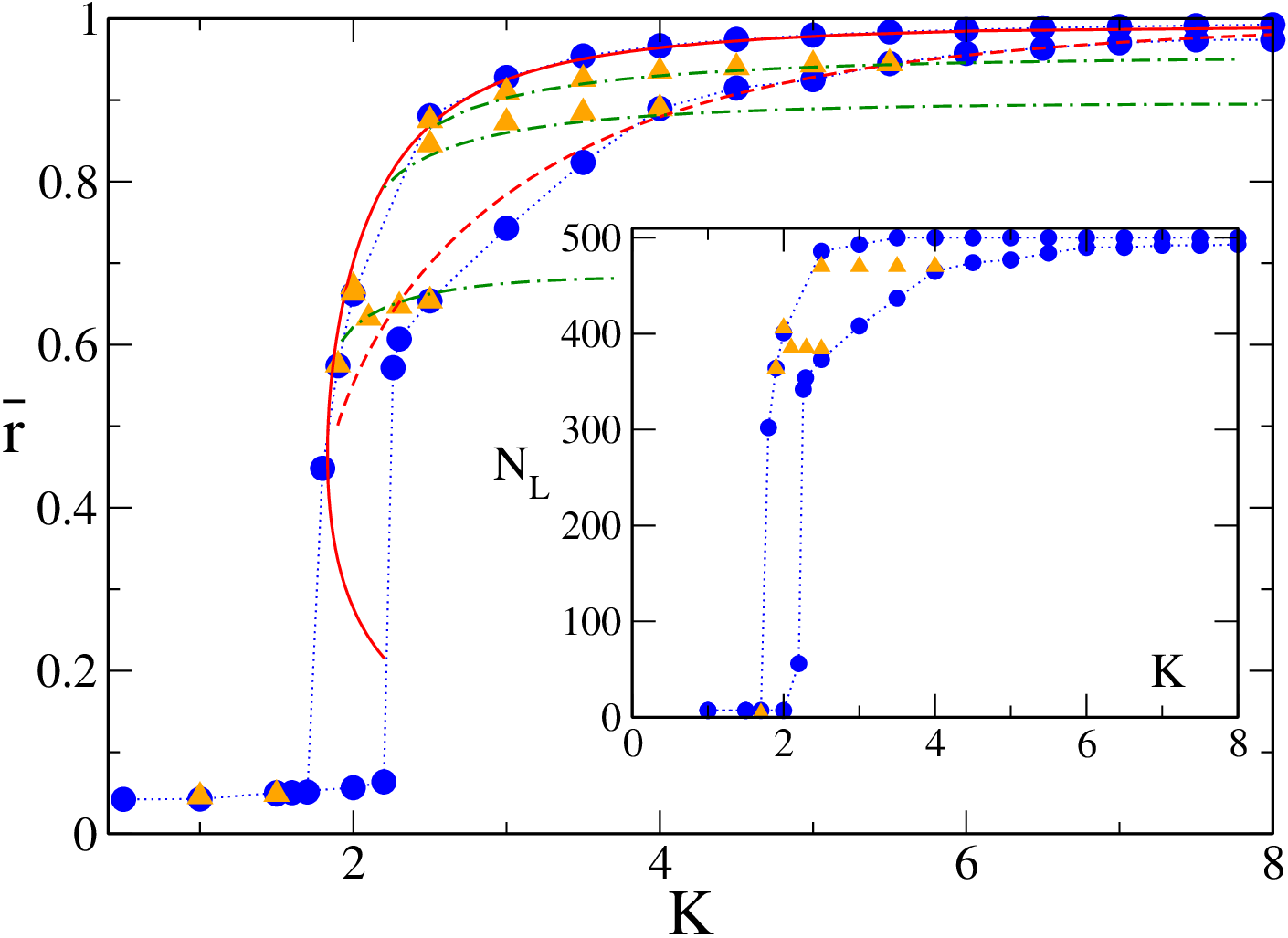}
\caption{(Color Online) Average order parameter ${\bar r}$ versus the coupling constant $K$
for a diluted network with 70\% of cutted links.  Mean field estimates: the dashed (solid) red curves refer 
to $r^I  = r^I_L + r^I_D$ ($r^{II}  = r^{II}_L + r^{II}_D$) as obtained by employing
Eqs. (\ref{rL}) and (\ref{rD}) following protocol I (protocol (II)); the (green) dot-dashed curves 
are the solutions $r^0(K,\Omega_0)$ of Eq. (\ref{r0}) for different $\Omega_0$ values.
The employed values from bottom to top are:  $\Omega_0= 2.05$, 1.69 and 1.10.
Numerical simulations: (blue) filled circles have been obtained by 
following protocol (I) and then (II) starting from $K=0$ until $K_M =20$ 
with steps $\Delta K=0.5$; (orange) filled triangles refer to simulations 
performed by starting from a final configuration obtained during protocol (I) and by decreasing 
the coupling from such initial configurations. The insets display $N_L$ vs $K$ for the numerical 
simulations reported in the main figures.
The numerical data refer to $m=2$, $N=500$, $N_c=150$, $T_R=5000$, and $T_W=200$.
\label{hysteresis_diluita}}
\end{figure}

Therefore, from this first analysis it emerges that the diluted or fully coupled systems,
whenever the coupling is properly rescaled with the in-degree, as in Eq.~\ref{eqPRL},
display the same phase diagram in the $(\bar r,K)$-plane even for very large dilution. 
In the following we will examine if the mean-field results obtained by following the TLO 
approach still apply to the diluted system. 
The comparison reported in figure Fig.~\ref{hysteresis_diluita} confirms the good agreement 
between the numerical results obtained
for a quite diluted system (namely, with 70 \% of broken links) and the mean-field predictions
 (\ref{rL}) and (\ref{rD}). Furthermore, the data reported in Fig.~\ref{hysteresis_diluita}
show that also in the diluted case all the states between the synchronization curves obtained following
protocol (I) and protocol (II) are reachable and \red{numerically stable}, analogously  
to what shown in Subsect IV A for the fully coupled system. These states, displayed as orange filled triangles 
in Fig.~\ref{hysteresis_diluita}, are characterized by a cluster composed 
by a constant number $N_L$ of locked oscillators with frequencies smaller than a value $\Omega_M$.
The number of oscillators in the cluster $N_L$ remains constant by varying the coupling between the
two synchronization curves (I) and (II).
Finally, the generalized mean-field solution $r^0(K,\Omega_0)$ 
(see Eq. (\ref{r0})) is able, also in the diluted case, to well reproduce the numerically obtained 
paths connecting the synchronization curves (I) and (II) (see Fig.~\ref{hysteresis_diluita} and the inset).

\section{A realistic network: the italian high-voltage power grid}
\label{sec6}

In this Section, we examine if the previously reported
features of the synchronization transition persist 
in a somehow more realistic setup. As we mentioned in the introduction a highly simplified 
model for a power grid composed of generators and consumers, resembling a Kuramoto model with inertia,
can be obtained whenever the generator dynamics can be expressed in terms
of the so-called swing equation~\cite{salam1984,filatrella2008}. The self-synchronization
emerging in this model has been recently object of investigation for different
network topologies~\cite{rohden2012,fortuna2012,rohden2013}. In this paper we will concentrate on
the Italian high-voltage (380 kV) power grid (Sardinia excluded), which is composed of 
$N=127$ nodes, divided in 34 sources (hydroelectric and thermal power plants)
and 93 consumers, connected by 342 links~\cite{fortuna2012}. This network is
characterized by a quite low average connectivity $\langle N_c \rangle= 2.865$,
due to the geographical distributions of the nodes along Italy~\cite{mappa}.

In this extremely simplified picture, each node can be described by its phase 
$\phi_i (t) = \omega_{\rm AC} t + \theta_i(t)$,
where $\omega_{\rm AC} = 2 \pi \times 50$ Hz or $2 \pi \times 60$ Hz is the standard AC
frequency and $\theta_i$ represents the phase deviation of the node $i$ 
from the uniform rotation at frequency $\omega_{\rm AC}$. Furthermore,
\red{the equation of motion  for each node is assumed to be the same for consumers and generators;
these are distiguished by the sign of a quantity $P_i$  associated each node: 
a positive (negative) $P_i$ corresponds to generated (consumed) power.}
By employing the conservation of energy and by assuming that the grid operates
in proximity of the AC frequency (i.e. $|\dot \theta| << \omega_{\rm AC}$)
and that the rate at which the energy is stored (in the kinetic term) is much
smaller than the rate at which is dissipated, the evolution equations for the phase
deviations take the following expression~\cite{filatrella2008}, 
\begin{equation}
\label{grid}
 \ddot{\theta}_i = \alpha \left[ - \dot{\theta}_i+ P_i + K \sum_j C_{i,j} \sin(\theta_j-\theta_i) \right]
\qquad .
\end{equation}
To maintain a parallel with the previously studied model (\ref{eqPRL}), we have multiplied
the left-hand side by a term $\alpha$, which in (\ref{grid}) represents the dissipations 
in the grid, while in (\ref{eqPRL}) corresponds to the inverse of the mass.
The parameter $\alpha \times K$ now represents the maximal power which can be 
transmitted between two connected nodes. 
More details on the model are reported in~\cite{salam1984,filatrella2008,rohden2013}.
It is important to stress that in order to have a stable, fully locked state, as
possible solution of (\ref{grid}), it is necessary that the sum of the generated power
equal the sum of the consumed power. Thus, by assuming that all the generators are 
identical as well as all the consumers, the distribution of the $P_i$
is made of two $\delta$-function located at $P_i = -C$ and $P_i = + G$. In
our simulations we have set $C=1.0$, $G=2.7353$ and $\alpha = 1/6$.  
This set-up corresponds to a Kuramoto model with inertia with a bimodal
distribution of the frequencies.
 
\begin{figure}
\includegraphics[angle=0,width=8.cm]{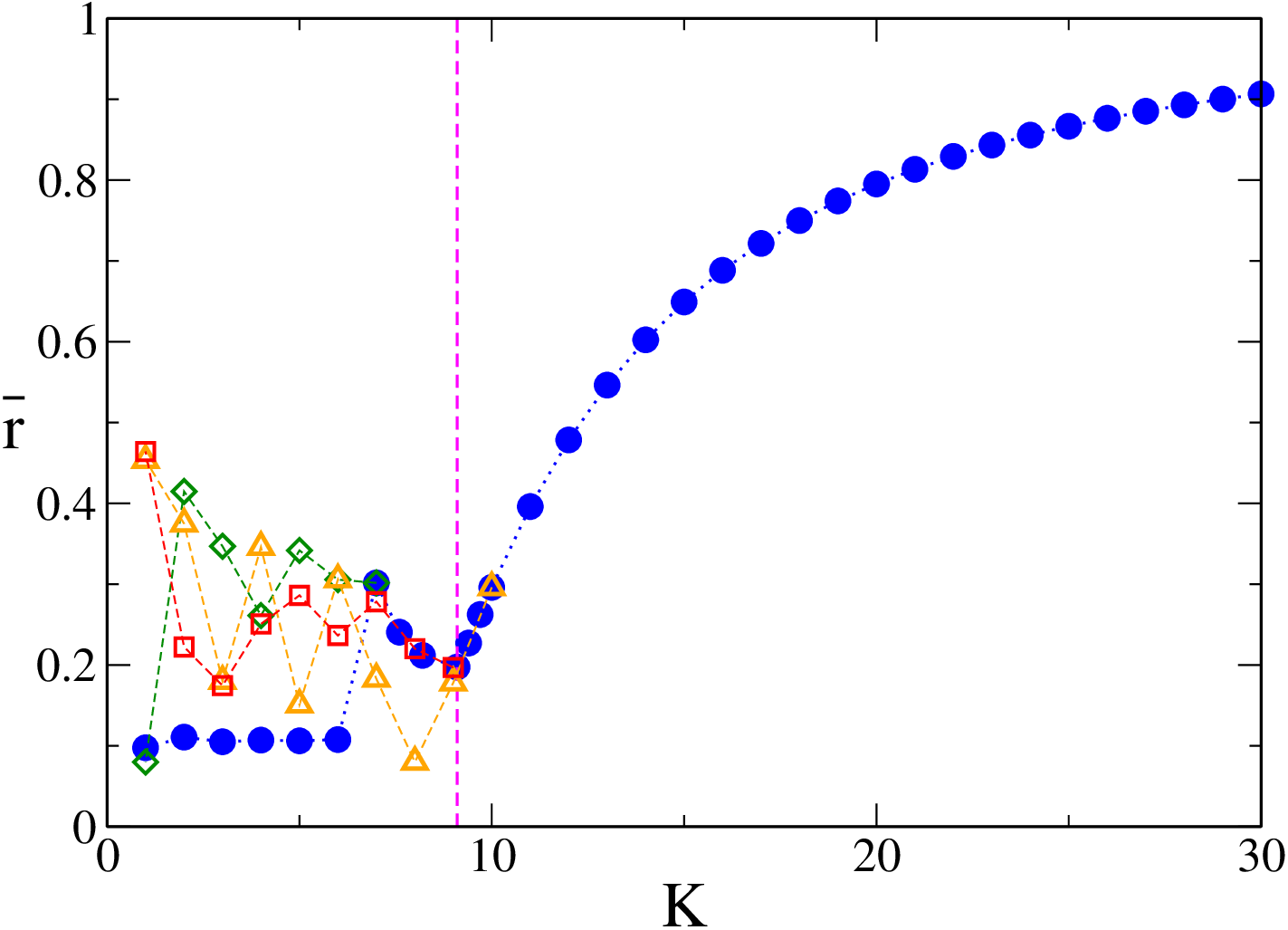}
\caption{(Color Online)  Average order parameter $\bar r$ versus the parameter $K$ for 
the Italian high-voltage power grid network. 
The (blue) circles data have been obtained by following
protocol  (I) from $K=1$ up to $K_M=40$ with $\Delta K = 1$. The other
symbols refer to simulations performed following protocol (II) 
starting from different intial coupling $K_I$ down to $K=1$, namely
(orange) triangles $K_I =10$, (red) squares $K_I=9$ and (green) diamond
$K_I=7$. The dashed vertical (magenta) line idicates the value $K=9$. The reported
data have been obtained by averaging the order parameter over 
a time window $T_W=5,000$, after discarding an initial transient time $T_R \simeq 60,000$.
\red{The numerical data refer to $\alpha=1/6$, N=127, $<N_c>=2.865$}.
\label{aver_italia}}
\end{figure}

As a first analysis we have performed simulations with protocol (I) for the model
(\ref{grid}) by varying the parameter $K$ and we have measured the corresponding average order 
parameter $\bar r$. As shown in Fig.~\ref{aver_italia} the behaviour of $\bar r$
with $K$ is non-monotonic. For small $K$ the state is asynchronous with
$\bar r \simeq 1/\sqrt{N}$, then $\bar r$ shows an abrupt jump for $K \simeq 7$
to a finite value, then it decreases reaching a minimum
at $K \simeq 9$. For larger $K$ the order parameter increases steadily with
$K$ tending towards the fully synchronized regime.

\begin{figure}
\includegraphics[angle=0,width=12.cm]{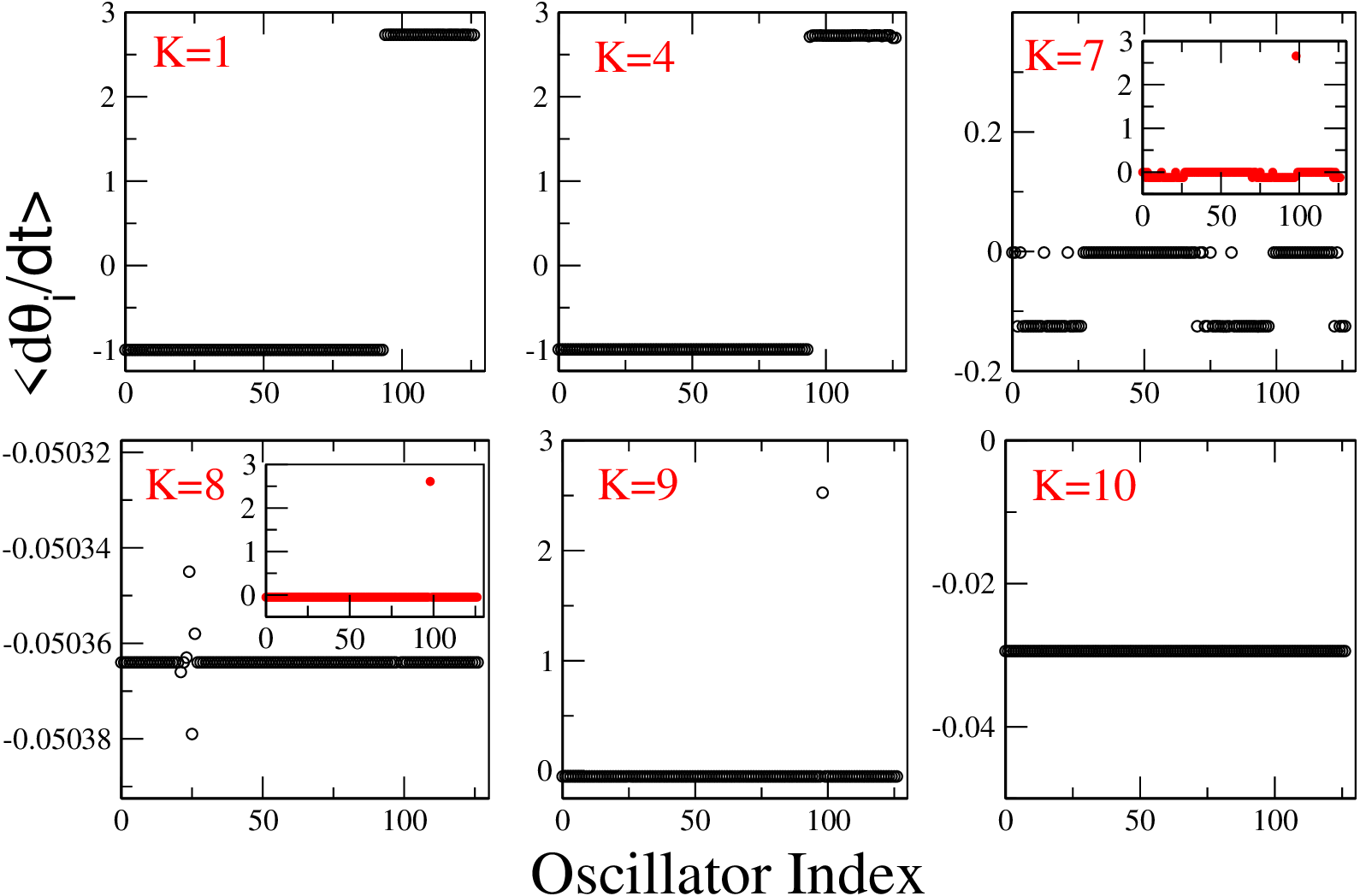}
\caption{(Color Online) 
Average phase velocity of each oscillator $\langle \dot \theta_i \rangle$ 
versus the oscillator index for different values of the coupling $K$. The oscillators have been reordered so that the first 
93 are consumers and the last 34 sources. The data have been obtained by employing protocol (I),
starting from zero coupling $K=0$ and with $\Delta K =1$. For each simulation an initial transient 
time $T_R \simeq 5,000$ has been discarded and the average is taken over a window $T_W = 5,000$.
\red{The numerical data refer to the same parameters as in Fig. \ref{aver_italia}}.
\label{clust_fwd}}
\end{figure}

\begin{figure}
\includegraphics[angle=0,width=8.cm]{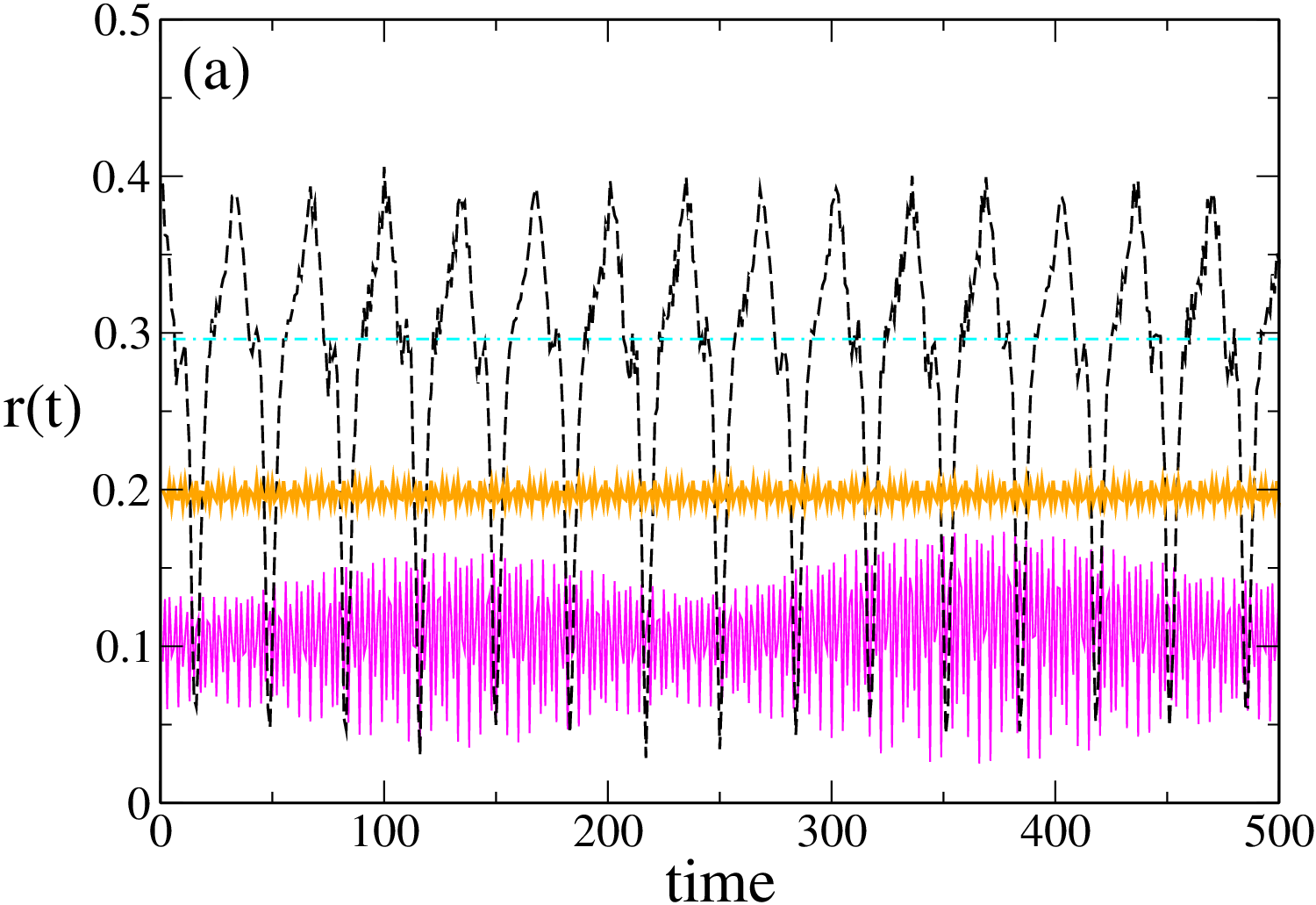}
\includegraphics[angle=0,width=8.cm]{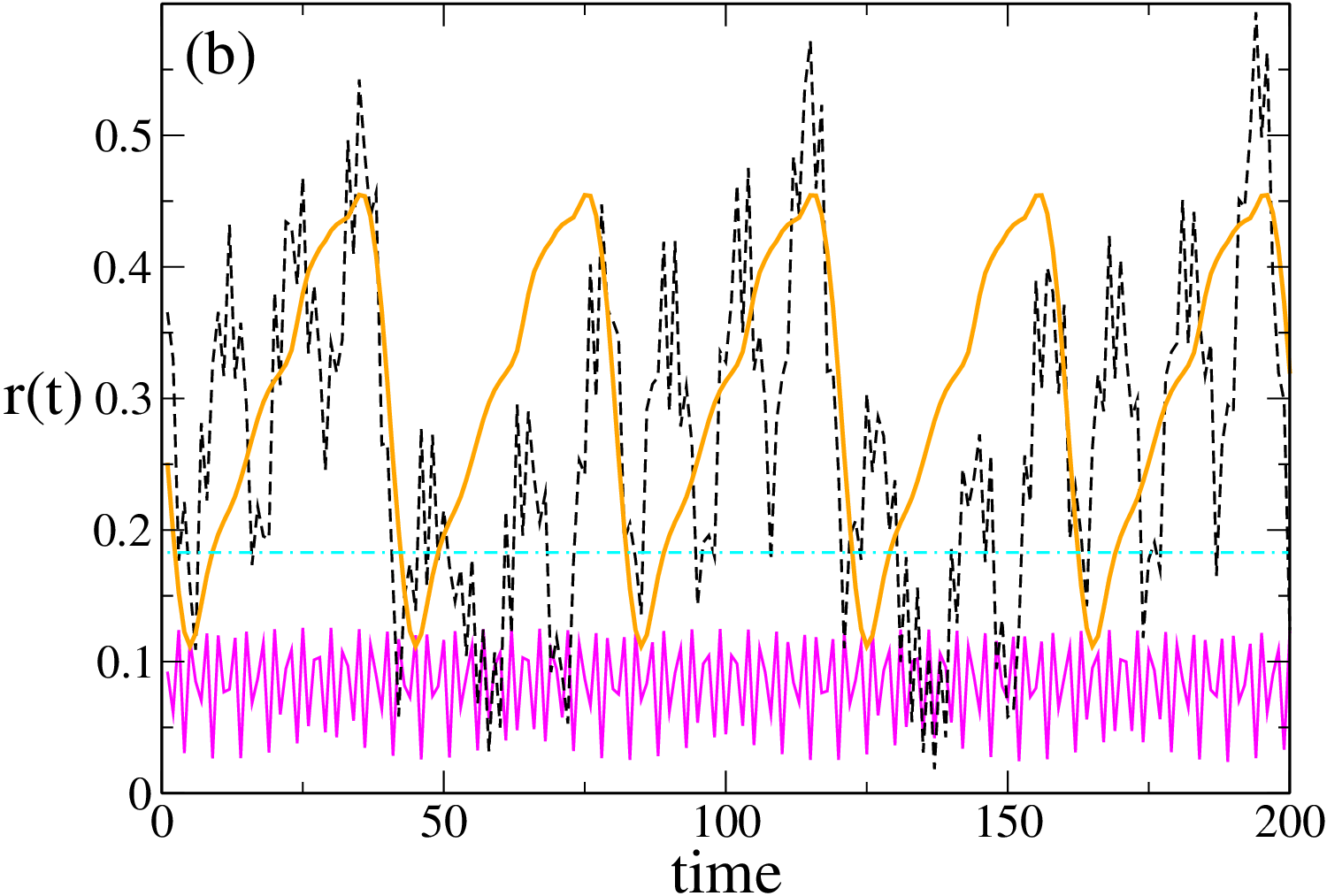}
\caption{(Color Online)
Order parameter $r(t)$ versus time for the Italian high-voltage power grid network for different 
values of the parameter $K$.\red{Panel (a): the dotted (magenta) curve corresponds 
to K=4; the dashed (black) line to K=7; the solid (orange) line to K=9; the dot-dashed (cyan) line to K=10.
The data have been obtained by employing protocol (I) and 
for each simulation  an initial transient time $T_R \simeq 5,000$ has been discarded.
Panel (b): the solid (magenta) curve corresponds to K=1; the dashed (black) line to K=4; the solid 
(orange) thick line to K=6; the  dot-dashed (cyan) line to K=7.}
The data have been obtained by employing protocol (II) and 
for each simulation  an initial transient time $T_R \simeq 50,000$ has been discarded.
\red{The simulations refer to the same parameters employed in Fig. \ref{aver_italia}}.
\label{r_timeItalia}}
\end{figure}

This behaviour can be understood by examining the average phase velocity of the oscillators
$\langle \dot \theta_i \rangle$. As shown in Fig.~\ref{clust_fwd},
for coupling $K < 7$ the system is splitted in 2 clusters: one composed by the sources
which oscillates with their proper frequency $G$ and the other one containing
the consumers, which rotates with average velocity $-C$. The oscillators in the 
two clusters rotate indipendently one from the other, therefore $\bar r \simeq 1/\sqrt{N}$.
For $K \simeq 7$ the oscillators get entrained (as shown in Fig.~\ref{clust_fwd})
and most of them are locked with almost zero average velocity, however a large 
part (50 over 127) form a secondary cluster of whirling oscillators with a velocity
$\langle \dot \theta \rangle \simeq -0.127$.
This secondary cluster has a geographical origin, since it includes power stations
and consumers located in the central part and south part of Italy, Sicily included.
The presence of this whirling cluster induces large oscillations
in the order parameter (see Fig.~\ref{r_timeItalia} (a)), reflecting almost
regular transitions from a desynchronized to a partially synchronized state. 
By increasing the coupling to $K=8$ the two clusters merge in an unique cluster with few scattered
oscillators, however the average velocity is small but not zero, 
namely $\langle \dot \theta \rangle \simeq - 0.05$
(as reported in Fig.~\ref{clust_fwd}). Therefore the average value of the order parameter $\bar r$
decreases with respect to $K=7$, where a large part of the oscillators was exactly locked.
Up to $K=9$, the really last node of the network, corresponding to one generator in Sicily 
connected with only one link to the rest of the Italian grid, 
still continues to oscillate indipendently from the other nodes, as shown in Fig.~\ref{clust_fwd}.
Above $K=9$ all the oscillators are finally locked in an
unique cluster and the increase in the coupling is reflected in a monotounous
increase in $\bar r$, similar to the one observed in standard Kuramoto models
(see Fig.~\ref{aver_italia}).

\begin{figure}
 \includegraphics[angle=0,width=12.cm]{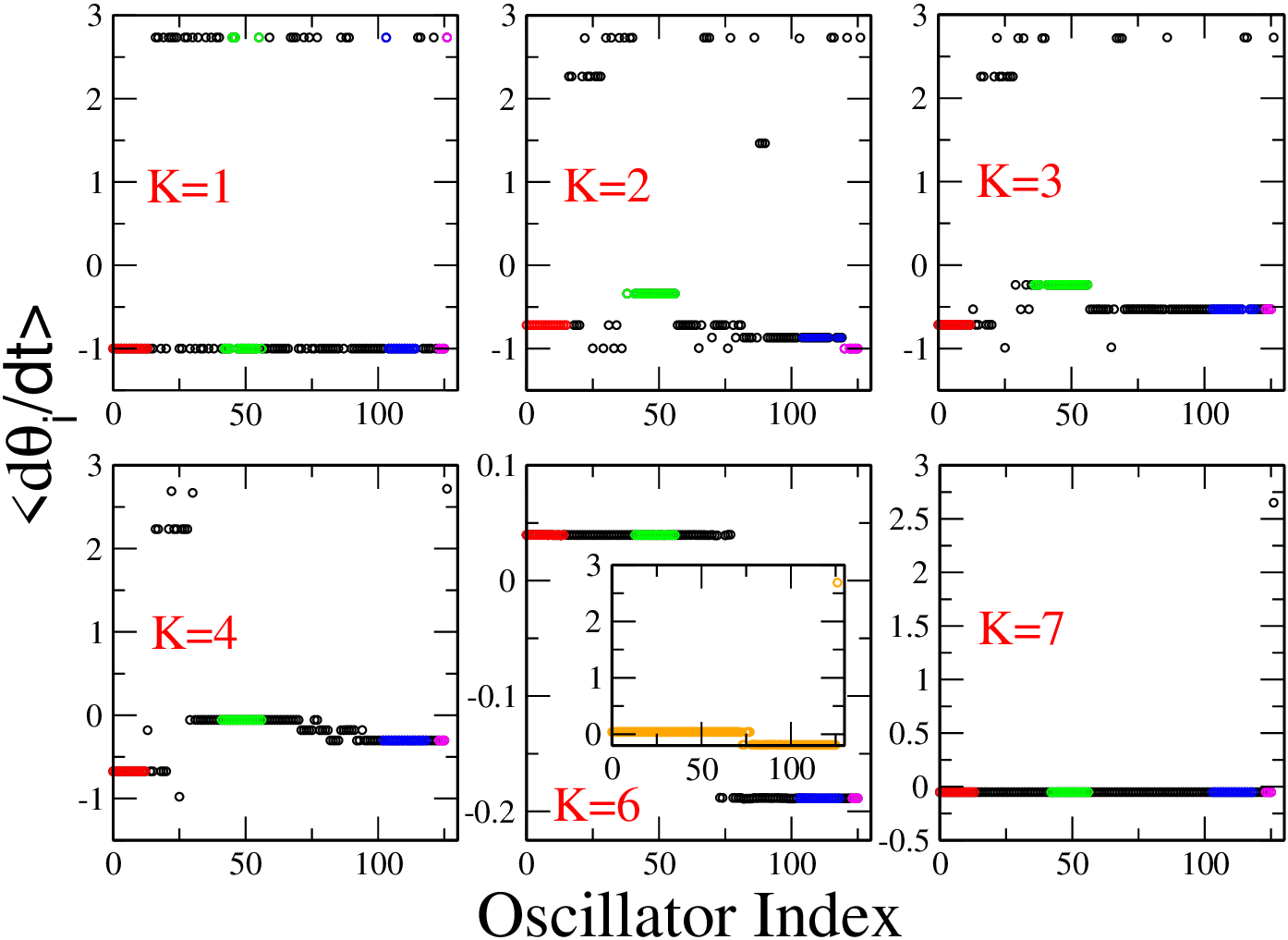}
 \caption{(Color Online) Average phase velocity of each oscillator $\langle \dot \theta_i \rangle$ 
versus the corresponding oscillator index, ordered following the geographical distribution
from north Italy to Sicily. The panels refer to different couplings.
The colored clusters indicate Italian regions which remains connected for 
all the considered simulations: red symbols refer to Piedmont and Liguria; 
green symbols to Veneto and Friuli Venetia Giulia; blue symbols to Campania and Apulia;
magenta symbols to Sicily. The data have been obtained by employing protocol (II) 
starting from $K_I = 12$ with $\Delta K =1$ down to $K=1$. 
For each simulation  an initial transient time $T_R \simeq 50,000$ has been discarded
and the averages performed over a window $T_W = 5000$.
\red{The numerical data refer to the same parameters as in Fig. \ref{aver_italia}}.
 \label{clust_back}}
\end{figure}

By applying protocol (II) we do not observe any hysteretic behaviour or
multistability down to $K=9$; instead for smaller coupling a quite intricated behaviour
is observable. As shown in Fig. \ref{clust_back} starting from $K_I=12$
and decreasing the coupling in steps of amplitude $\Delta K =1$, the system stays mainly in
one single cluster up to $K=7$, apart the last node of the network which
already detached from the network at some larger $K$. Indeed at $K=7$ the order parameter
has a constant value around 0.2 and no oscillations. 
As shown in Fig.\ref{r_timeItalia} (b), by decreasing the
coupling to $K=6$, wide oscillations emerge in $r(t)$ due to the fact that the
locked cluster has splitted in two clusters, the separation is
similar to the one reported for $K=7$ in Fig.~\ref{clust_fwd}. 
By further lowering $K$, several small whirling clusters appear
and the behaviour of $r(t)$ becomes seemingly irregular for $2 \le K \le 5$
as reported in Fig.\ref{r_timeItalia} (b).
An accurate analysis of the dynamics in terms of the maximal Lyapunov
exponent has revealed that the irregular oscillations in $r(t)$ reflect
quasi-periodic motions, since the measured maximal Lyapunov is always zero 
for the whole range of the considered couplings.
\red{The presence of the inertial term, together with an architecture
which favours a splitting based on the proximity of the oscillators,
lead to the formation of several whirling clusters characterized
by different average phase velocities.} The value of the order
parameter arises as a combination of these different contributions,
each corresponding to a different oscillatory frequency.
 The splitting in different clusters
is probably also at the origin of the multistability observed 
for $K < 7$: depending on the past history the grid splits in
clusters formed by different groups of oscillators and this gives
rise to different average values of the order parameter (see Fig.~\ref{aver_italia}).

\red{We have verified that the emergence of several whirling clusters, with
an associated quasi-periodic behaviour of the order parameter, is observable
also by considering an unimodal (Gaussian) distribution of the $P_i$. 
This confirms that the main ingredients at the origin of this phenomenon 
are the inertial term together with a short-range connectivity. 
Thus the bimodal distribution, here employed, seems not to be crucial
and it can only lead to an enhancement of such effect.}

\section{Conclusions}
\label{sec7}

We have studied the synchronization transition for a globally
coupled Kuramoto model with inertia for different system sizes and inertia values.
The transition from incoherent to coherent state is hysteretic 
for sufficiently large masses. \red{In particular, the upper value of the
coupling constant ($K_1^{c}$), for which an incoherent state is observable, 
increases with the system sizes for all the examined masses. The estimated finite size value 
$K_1^{c}$  has a non monotonic dependence on the mass $m$, exhibiting a maximum
at some intermediate value of $m$. However, all the data obtained for
different masses and sizes collapse onto an universal curve,
whenever the distance of $K_1^{c}$ with respect to its 
mean field value~\cite{gupta2014} is
reported as a function of the mass divided by $N^{1/5}$. On the
other hand, the coherent phase is attainable above a minimal critical coupling ($K_2^{c}$)
which exhibits a weak dependence on the system size and it saturates to
a constant asymptotic value for sufficiently large inertia values.}

\red{Furthermore, we have shown that clusters of locked oscillators 
of any size coexist within the hysteretic region. This region
is delimited by two curves in the plane individuared by the coupling and the 
average value of the order parameter. 
Each curve corresponds to the {\it synchronization} ({\it desynchronization})
profile obtained  starting from the fully desynchronized (synchronized) state.}
The original mean field theory developed 
by Tanaka, Lichtenberg, and Oishi in 1997~\cite{tanaka1997first,tanaka1997self} 
gives a reasonable estimate of both these limiting curves, while 
a generalization of such theory is capable to reproduce
all the possible synchronization/desynchronization hysteretic loops.
However, the TLO theory does not take into account the presence of clusters
composed by drifting oscillators emerging for sufficiently large
masses. The coexistence of these clusters with the cluster of locked
oscillator induces oscillatory behaviour in the order parameter. 

The properties of the hysteretic transition have been examined also
for random diluted network; the main properties of the transition
are not affected by the dilution up to extremely high values.
The transition appears to become continuous only when the 
number of links per node becomes of the order of few units.
By increasing the system size the transition to the continuous 
case (if any) shifts to smaller and smaller values of the connectivity.

In this paper we focused on Gaussian distribution of the natural frequencies,
however we have obtained similar results also for Lorentzian distributions. 
It would be however interesting to examine how the transition modifies
in presence of non-unimodal distributions for the natural frequencies,
like bimodal ones. Preliminary indications in this
direction can be obtained by the reported analysis of the self-synchronization 
process occurring in the Italian high-voltage power grid, when the generators
and consumers are mimicked in terms of a Kuramoto model
with inertia~\cite{filatrella2008}. In this case
the transition is largely non hysteretic, probably this
is due to the low value of the average connectivity in such a network. 
Coexistence of different states made of whirling and locked clusters, formed on regional basis, 
is observable only for electrical lines with a low value of the maximal transmissible power.
These states are characterized by quasi-periodic oscillations
in the order parameter due to the coexistence of several clusters of drifting oscillators.

A natural prosecution of the presented analysis would be the
study of the stability of the observed clusters of locked
and/or whirling oscillators \red{in presence of noise}.
In this respect, exact mean-field results have been reported recently
for fully coupled phase rotors  with inertia and additive noise~\cite{gupta2014,komarov2014}. 
However, the emergence of clusters in such systems has been not yet
addressed neither on a theoretical basis nor via direct simulations.

\begin{acknowledgments}
We acknowledge useful discussions with  J. Almendral,
M. B\"ar, I. Leyva, A. Pikovsky, J. Restrepo, S. Ruffo, and  I Sendi\~na-Nadal,
and we thank M. Frasca for providing the connectivity matrix
relative to the Italian grid. Financial support has been 
given by the Italian Ministry of University and Research 
within the project CRISIS LAB PNR 2011-2013. SO and AT thank 
the German Science Foundation DFG, within the framework of 
SFB 910 "Control of self-organizing nonlinear systems``,
for the kind hospitality offered during 2012 and 2013
at Physikalisch-Technische Bundesanstalt in Berlin.
\end{acknowledgments}


%

\end{document}